\pdfoutput=1
\documentclass[traditabstract]{aa} 
\usepackage{graphicx,natbib}
\usepackage{longtable,lscape,supertabular}
\usepackage{txfonts}

\begin{document}
\title{Unveiling shocks in planetary nebulae} 


\author{Mart\'{\i}n A.\ Guerrero\inst{1}, 
         Jes\'us A.\ Toal\'a\inst{1}, 
         Juan J.\ Medina\inst{1}, 
         Valentina Luridiana\inst{2,3}, 
         Luis F.\ Miranda\inst{4,5}, 
         Angels Riera\inst{6,7}, 
        \and
         Pablo F.\ Vel\'azquez\inst{8}
        }


\institute{
Instituto de Astrof\'{\i}sica de Andaluc\'{\i}a, IAA-CSIC, 
Glorieta de la Astronom\'\i a s/n, E-18008, Granada, Spain;  
\and
Instituto de Astrof\'\i sica de Canarias, IAC, 
V\'\i a L\'actea s/n, E-38205, La Laguna, Spain; 
\and
Departamento de Astrof\'\i sica, Universidad de La Laguna, E-38200 La 
Laguna, Tenerife, Spain; 
\and
Departamento de F\'\i sica Aplicada, Facultade de Ciencias, 
Campus Lagoas-Marcosende s/n, Universidade de Vigo, E-36310, Vigo, Spain;  
\and
Consejo Superior de Investigaciones Cient\'\i ficas (CSIC), 
c/ Serrano 117, E-28006 Madrid, Spain; 
\and
Departament de F\'\i sica i Enginyeria Nuclear, EUETIB, 
Universitat Polit\`ecnica de Catalunya, Comte d'Urgell 187, 
08036, Barcelona, Spain; 
\and
Departament d'Astronomia i Meteorologia, Universitat de 
Barcelona, 
Av. Diagonal 647, 08028 Barcelona, Spain;  
\and
Instituto de Ciencias Nucleares, UNAM, 
Ciudad Universitaria, 04510, Mexico City, Mexico. \\
\email{mar@iaa.es}
}

\date{Received ; accepted }

 
\abstract{ 
The propagation of a shock wave into a medium is expected to heat the 
material beyond the shock, producing noticeable effects in intensity 
line ratios such as [O~{\sc iii}]/H$\alpha$.  
To investigate the occurrence of shocks in planetary nebulae (PNe), we have 
used all narrow-band [O~{\sc iii}] and H$\alpha$ images of PNe available in 
the \emph{HST} archive to build their [O~{\sc iii}]/H$\alpha$ ratio maps and 
to search for regions where this ratio is enhanced.  
Regions with enhanced [O~{\sc iii}]/H$\alpha$ emission ratio can 
be ascribed to two different types of morphological structures: 
bow-shock structures produced by fast collimated outflows and thin 
skins enveloping expanding nebular shells.  
Both collimated outflows and expanding shells are therefore confirmed 
to generate shocks in PNe.  
We also find regions with depressed values of the [O~{\sc iii}]/H$\alpha$ 
ratio which are found mostly around density bounded PNe, where the 
local contribution of [N~{\sc ii}] emission into the F656N H$\alpha$ 
filter cannot be neglected.
}


\authorrunning{Guerrero et al}
\titlerunning{Unveiling Shocks in PNe}
\maketitle
%

\section{Introduction}

The increase in the local pressure of the material in a region makes it to 
expand, producing a disturbance in the surroundings that may drive a shock 
front.  
Shocks are ubiquitous in the universe whenever supernova explosions, 
photoionized gas, stellar winds, and collimated outflows drive 
shock waves in the circumstellar or interstellar media.

Shocks of fast moving outflows with material in the surrounding 
medium should produce appreciable effects.  
The most obvious is probably the formation of a bow-shock structure, as 
those associated with jet-like outflows in Herbig-Haro objects such as 
the archetypical HH\,34 and HH\,111 
\citep{Reipurth_etal1997,Reipurth_etal2002,Raga_etal02}.  
Planetary nebulae (PNe) also exhibit fast collimated outflows 
whose occurrence is found at very early stages of the nebular 
formation, including the late stages of the asymptotic giant 
branch (AGB) and the proto-PN phases. 
These collimated outflows can produce notable bow-shock features, 
as those detected in IC\,4593 \citep{Corradi_etal1997}.

On microscopic scales, a shock wave is an irreversible process, where 
entropy is generated as ordered kinetic energy is dissipated into heat.  
Therefore, a fast outflow propagating into a low density medium 
drives a forward shock which produces an increase of the electron 
temperature, $T_{\rm e}$.  
The increase in temperature would enhance the emission in the [O~{\sc iii}] 
$\lambda$5007 \AA\ line, which is very sensitive to $T_{\rm e}$, whereas the 
low density would reduce the emissivity of the H$\alpha$ line, which is more 
sensitive to the electron density, $N_{\rm e}$.  
Consequently, it is expected a region in front of the bow-shock at the 
tip of a collimated outflow where the [O~{\sc iii}]/H$\alpha$ ratio is 
significantly enhanced.  
Such caps of enhanced [O~{\sc iii}]/H$\alpha$ ratio have been revealed 
by \emph{HST} images of IC\,4634, a PN characterized by the presence 
of fast collimated outflows resulting in a point-symmetric morphology 
and clear bow-shock structures \citep{Guerrero_etal2008}.


A dense, fast shell expanding into a tenuous, static medium can also 
propagate shocks into the surroundings.  
This effect is observed in bubbles blown by the wind of massive stars, 
such as S\,308, the wind-blown bubble around the Wolf-Rayet star HD\,50896, 
whose shell 
generates a forward shock that 
produces a notable offset between the [O~{\sc iii}] and H$\alpha$ emissions 
\citep{Gruendl_etal2000}.  
Another example is provided by NGC\,7635, a.k.a.\ the Bubble Nebula, a 
bubble blown by the wind of the massive O6.5\,IIIf star BD+60\degr2522, 
where the offset between these emissions is $\approx$3$\times$10$^{15}$ 
cm \citep{Moore_etal2002}.  
The expansion of the different shells of PNe is also expected to 
generate shocks: 
one at the leading edge of the outer shell propagating into the 
unperturbed AGB wind and another at the leading edge of the bright 
rim expanding into the outer shell \citep[e.g., ][]{Petal04}.  
\citet{Balick2004} found a thin skin of enhanced 
[O~{\sc iii}]/H$\alpha$ enveloping NGC\,6543 in 
\emph{HST} images of this nebula.  
The origin of this skin of enhanced [O~{\sc iii}]/H$\alpha$ remained 
uncertain, but an observational artefact could be ruled out.



Fast collimated outflows are acknowledged to play an essential role in the 
nebular shaping in early stages of the PN formation \citep{ST98}.  
In particular, the expansion of bow-shock features has drawn the 
attention as a mechanism for the formation of bipolar and multi-polar 
PNe \citep[see the review by][]{BF02}.  
Similarly, the expansion of the different nebular shells is critical 
for the evolution of PNe \citep{Villaver_etal2002,Petal04}.  
Both the shocks produced by fast collimated outflows and those 
produced by the shell expansion may have important dynamical 
effects in PNe, contributing decisively to the nebular shaping, 
evolution, and excitation.  
%
%
%

Inspired by the results obtained for IC\,4634 and NGC\,6543, we have used 
the \emph{HST} [O~{\sc iii}] and H$\alpha$ images of PNe available in the 
Mikulski Archive for Space Telescopes (MAST) to investigate the occurrence 
of skins of bright [O~{\sc iii}]/H$\alpha$ in PNe and its relationship 
with fast collimated outflows and expanding shells.  
The description of the search and analysis of \emph{HST} images is 
presented in Sect.~2.  
Attending to their [O~{\sc iii}]/H$\alpha$ ratio maps (Sect.~3), the PNe 
in this sample can be divided in four different types of which two of 
them show regions of enhanced [O~{\sc iii}]/H$\alpha$ emission associated 
with bow-shocks of collimated outflows and with expanding nebular shells, 
respectively.  
The results are further discussed in Sect.~4 and a short summary is 
given in Sect.~5.

\begin{figure*}[htbp]
\centering
\setlength{\fboxsep}{0.1pt}
\includegraphics[width=0.9\linewidth]{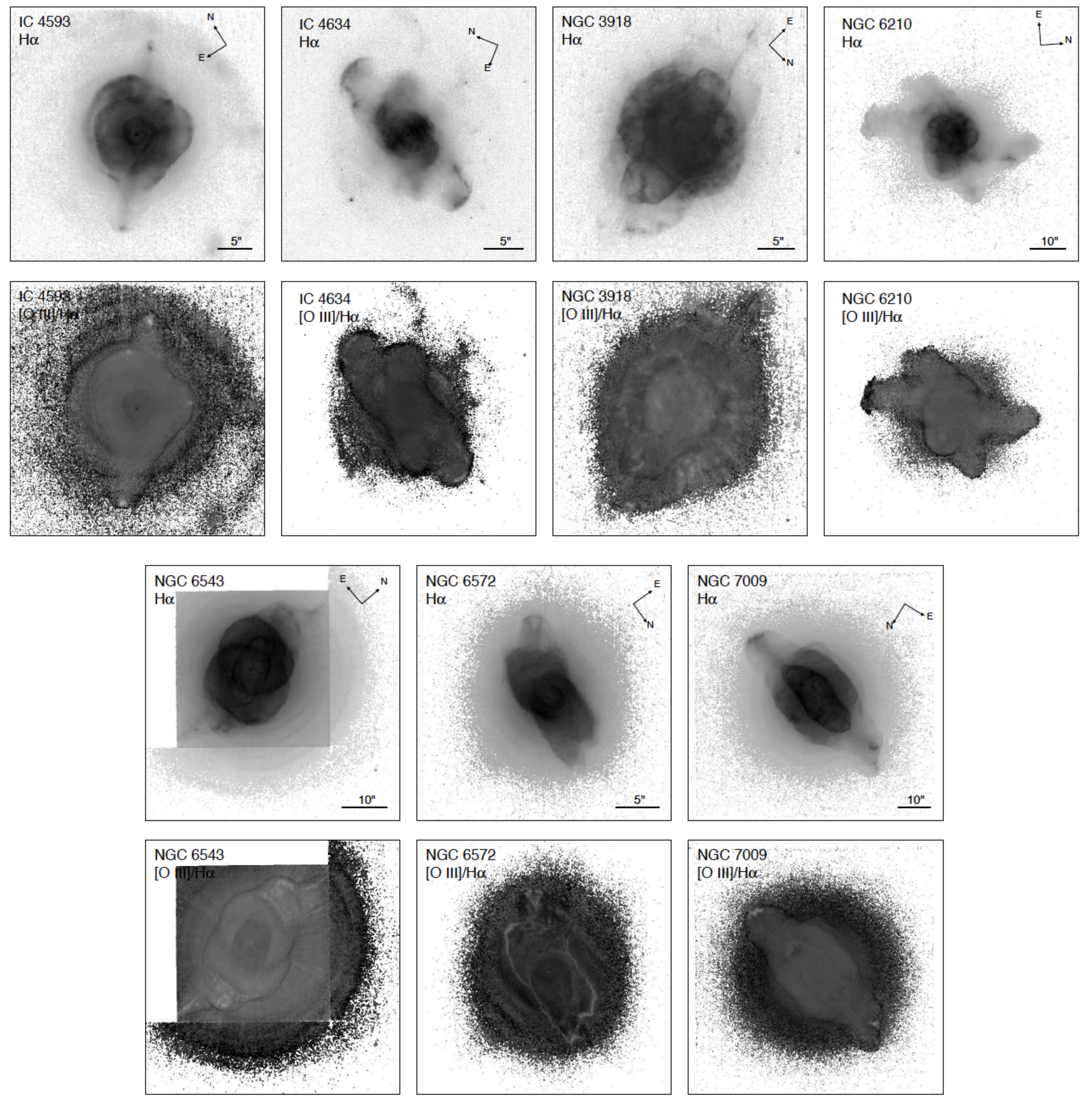}
\caption{
H$\alpha$ and {[O~{\sc iii}]/H$\alpha$} ratio maps of the type A PNe 
IC\,4593, IC\,4634, NGC\,3918, NGC\,6210, NGC\,6543, NGC\,6572, and 
NGC\,7009.  
In the {[O~{\sc iii}]/H$\alpha$} ratio maps, regions of relatively 
bright H$\alpha$ emission appear bright (white), whereas regions of 
strong [O~{\sc iii}] emission relative to H$\alpha$ are shown dark 
(black).  
}  
\label{fig_typeA}
\end{figure*}

\section{H$\alpha$ and [O~{\sc iii}] Images of PNe in the HST Archive}

We searched MAST for \emph{HST} WFPC2 or WFC3 coeval H$\alpha$ and 
[O~{\sc iii}] images of PNe available by March 2013.  
This search yielded H$\alpha$ and [O~{\sc iii}] images for 103 PNe obtained 
through the F656N and F502N filters, respectively (Table~\ref{tab1}).  
Since the adjacent [N~{\sc ii}] $\lambda\lambda$6548,6584 \AA\ emission lines 
can be included in the bandpass of the F656N H$\alpha$ narrow-band filter, the 
potential contamination of the emission from these low-excitation lines on the 
H$\alpha$ image needs to be discussed into further detail.

The bandpass of the WFC3 F656N H$\alpha$ filter only includes the 
contribution from the [N~{\sc ii}] $\lambda$6548 \AA\ emission 
line with a filter transmission at its wavelength which is 
$\sim$5\% that at the wavelength of the targeted H$\alpha$ line 
\citep{ODell_etal2013}.  
As for the WFPC2 F656N H$\alpha$ filter, its bandpass is broader 
\citep{Lim_etal2010} and includes contributions of both [N~{\sc ii}] 
lines, with filter transmissions relative to that at the wavelength 
of the H$\alpha$ line of $\sim$33\% for the [N~{\sc ii}] $\lambda$6548 
\AA\ component and $\sim$5\% for the [N~{\sc ii}] $\lambda$6584 \AA\ 
line.  
Therefore, we can expect a 10\% contamination by [N~{\sc ii}] emission in the 
WFC3 F656N filter when the [N~{\sc ii}] $\lambda$6584 \AA\ line is 6 times 
brighter than the H$\alpha$ line, but only $\sim$2/3 as bright for the WFPC2 
F656N filter.

\begin{figure*}[htbp]
\centering
\setlength{\fboxsep}{0.1pt}
\includegraphics[width=0.9\linewidth]{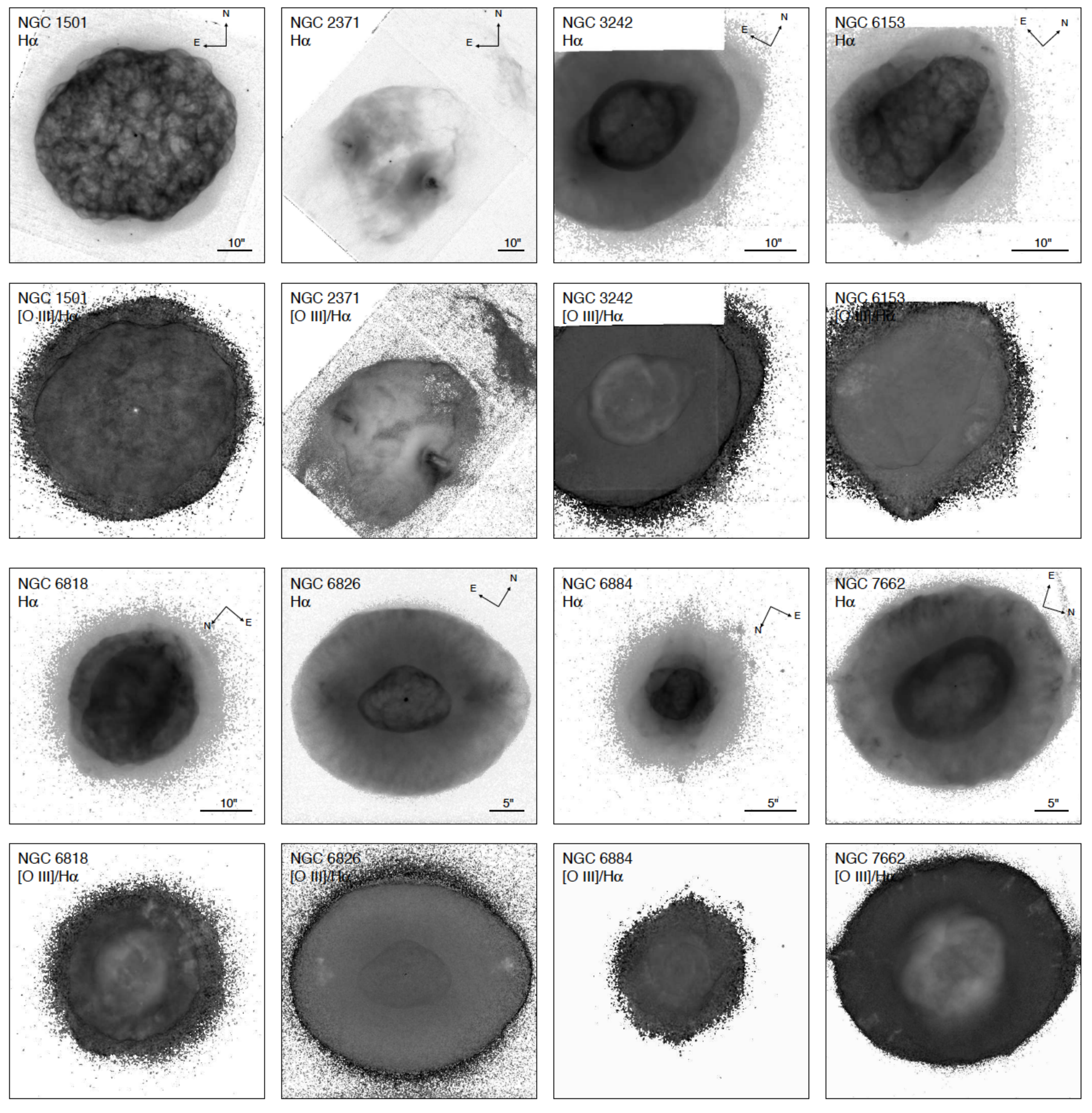}
\begin{center}
\caption{
Same as Fig.~\ref{fig_typeA} for the type B PNe 
NGC\,1501, NGC\,2371-2, NGC\,3242, NGC\,6153, NGC\,6818, 
NGC\,6826, NGC\,6884, and NGC\,7662.  
}
\label{fig_typeB}
\end{center}
\end{figure*}

The [N~{\sc ii}] to H$\alpha$ line ratio may be large for 
low-excitation PNe with bright [N~{\sc ii}] emission lines. 
To preserve our sample from sources whose H$\alpha$ image may be significantly 
contaminated by the contribution of [N~{\sc ii}] emission, hereafter we will 
exclude from our analysis those objects observed by the WFPC2 instrument with 
a [N~{\sc ii}] $\lambda$6584 \AA\ to H$\alpha$ line ratio greater than unity.  
This includes H\,1-9, H\,2-15, Hb\,5, IC\,4406, K\,4-55, M\,2-14, 
Mz\,3, NGC\,2438, NGC\,2440, NGC\,2818, NGC\,3132, and NGC\,6537.  
The source NGC\,6302, observed with the WFC3 instrument, will be further 
excluded because its extremely high [N~{\sc ii}] $\lambda$6584 \AA\ to 
H$\alpha$ line ratio.  
The [O~{\sc iii}]/H$\alpha$ ratio maps of sources with a 
[N~{\sc ii}] $\lambda$6584 \AA\ to H$\alpha$ line ratio 
close to unity will be carefully scrutinized to check 
for features associated to [N~{\sc ii}] contamination.  
We note that the (H$\alpha$+[N~{\sc ii}])/[O~{\sc iii}] ratio maps of PNe 
notably enhance the [N~{\sc ii}] emission of low-ionization small-scale 
structures \citep{Corradi_etal96}.  
These features are easy to identify.

The images were downloaded from the \emph{HST} archive and reduced using 
the standard pipeline procedure.  
For each PN, individual frames of the same epoch were combined to 
remove the cosmic rays using IRAF\footnote{
IRAF is distributed by the National Optical Astronomy Observatories 
which is operated by the Association of Universities for Research in 
Astronomy, Inc.\ (AURA) under cooperative agreement with the National 
Science Foundation.  
}
routines.  
The resulting H$\alpha$ and [O~{\sc iii}] images were subsequently 
used to generate [O~{\sc iii}] to H$\alpha$ ratio maps.  
Before producing these ratios, we checked the relative alignment of the 
images in the different filters.  
The WFC3 images were notably well aligned, but we noticed a consistent shift 
of 0\farcs1--0\farcs3 between the WFPC2 H$\alpha$ and [O~{\sc iii}] images 
caused most likely by the different light path across the filters.  
Accordingly, the images in the different filters were aligned using stars in 
the field of view (including the central stars when available) and registered 
on the same grid.  
The accuracy of this procedure was usually better than 10\% of a 
pixel, i.e., $\lesssim$0\farcs005.

\begin{figure*}[htbp]
\begin{center}
\setlength{\fboxsep}{0.1pt}
\includegraphics[width=0.9\linewidth]{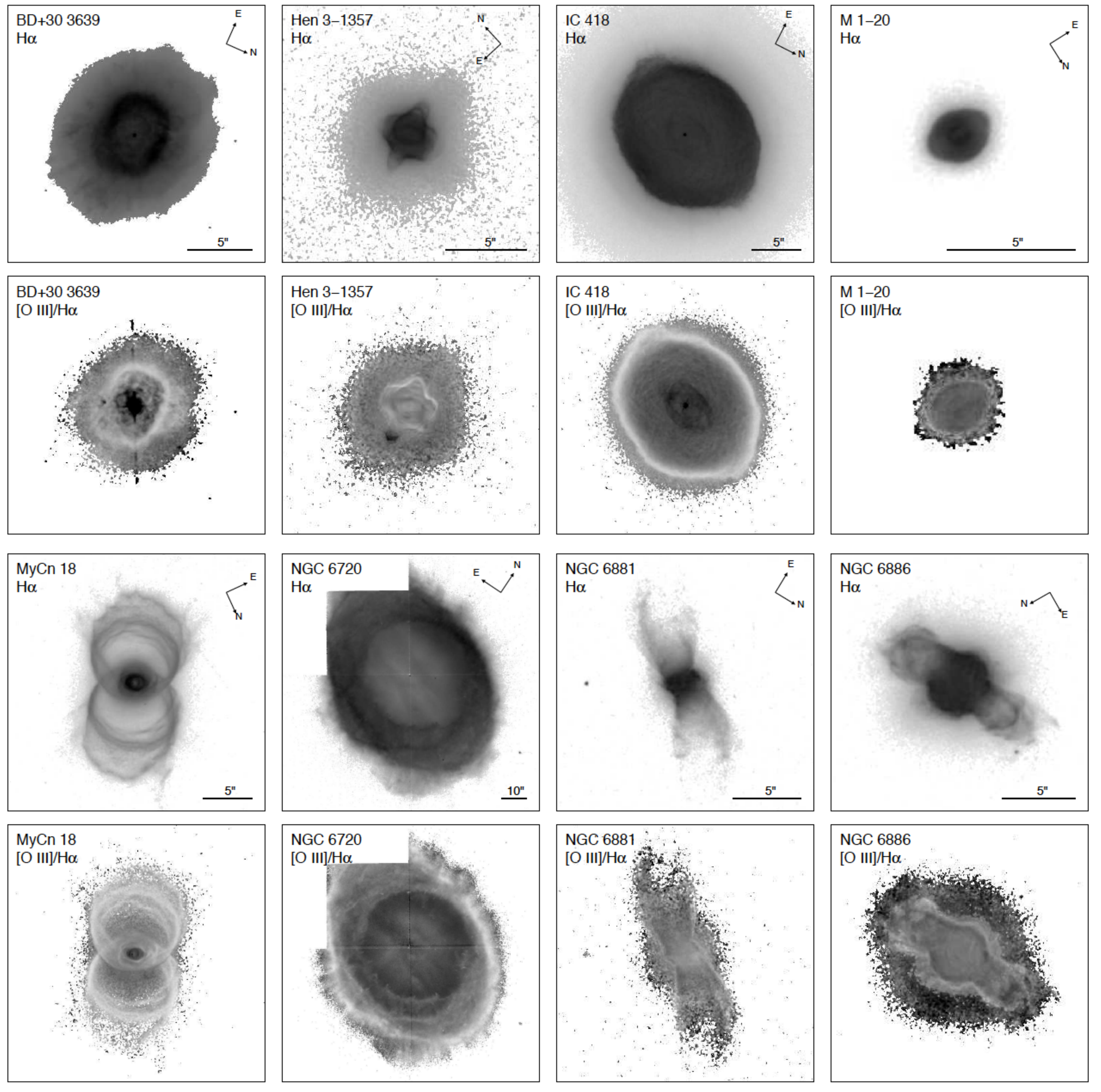}
\caption{
Same as Fig.~\ref{fig_typeA} for the type C PNe.  
}
\label{fig_typeC}
\end{center}
\end{figure*}

The [O~{\sc iii}]/H$\alpha$ ratio maps were examined to investigate the 
relative positions of the [O~{\sc iii}] and H$\alpha$ emissions and to 
search for regions of enhanced values of the [O~{\sc iii}]/H$\alpha$ 
ratio.  
The images of 18 PNe were found not suitable for this analysis: 
NGC\,6853 (the Dumbbell Nebula) and NGC\,7293 (the Helix Nebula) are much 
larger than the WFPC2 field of view, 
SaSt\,1-1 and Th\,4-1 are point sources, 
Hen\,2-436, NGC\,6833, and PC\,11 are compact PNe whose emission is not 
properly resolved, and 
H\,1-43, H\,1-55, H\,2-25, H\,2-26, Hen\,2-262, JaFu\,2, K\,3-76, KFL\,7, 
Th\,3-6, Th\,3-14, and Wray\,16-282 are weak in [O~{\sc iii}], resulting 
in poor quality [O~{\sc iii}]/H$\alpha$ ratio maps. 
All these 18 PNe have been discarded from subsequent analysis.

\begin{figure*}[htbp]
\begin{center}
\setlength{\fboxsep}{0.1pt}
\includegraphics[width=0.9\linewidth]{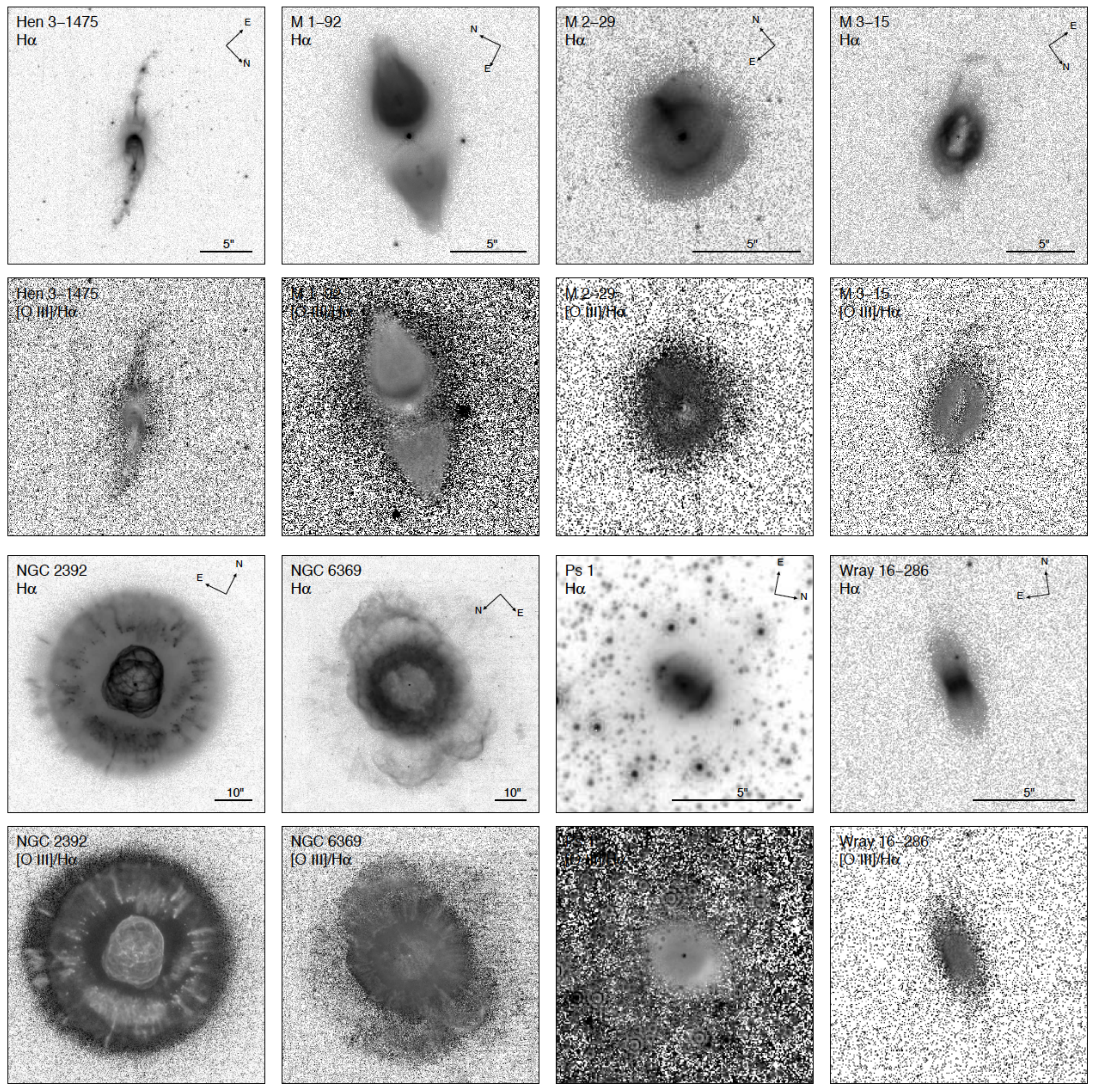}
\caption{
Same as Fig.~\ref{fig_typeA} for a subsample of type D PNe. 
}
\label{fig_typeD}
\end{center}
\end{figure*}


\section{PN Classification Based on the [O~{\sc iii}]/H$\alpha$ Ratio}

The remaining 72 PNe in our sample present [O~{\sc iii}]/H$\alpha$ 
ratio maps suitable for its analysis.  
The H$\alpha$ and [O~{\sc iii}]/H$\alpha$ ratio images of a representative 
number of PNe with adequate [O~{\sc iii}]/H$\alpha$ ratio maps are presented 
in Figs.~\ref{fig_typeA} through \ref{fig_typeD}.  
Grey-scales in the [O~{\sc iii}]/H$\alpha$ ratio have been chosen to highlight 
the locations where the [O~{\sc iii}] and H$\alpha$ emissions show spatial 
differences.  
A close inspection of the H$\alpha$ and [O~{\sc iii}]/H$\alpha$ ratio 
images of these PNe has allowed us to classify them into four different 
types: 
\begin{itemize}
\vspace*{-0.20cm}
\item {\bf Type A} \\
PNe that present caps of enhanced {[O~{\sc iii}]/H$\alpha$} ratios 
at the tip of jet-like features.  
These regions of bright [O~{\sc iii}] emission can be interpreted as 
bow-shock features produced by fast collimated outflows.  
The H$\alpha$ images and {[O~{\sc iii}]/H$\alpha$} ratio maps of the type 
A PNe IC\,4593, IC\,4634, NGC\,3918, NGC\,6210, NGC\,6543, NGC\,6572, and 
NGC\,7009 are shown in Fig.~\ref{fig_typeA}. 
\item {\bf Type B} \\
PNe surrounded by skins of enhanced {[O~{\sc iii}]/H$\alpha$} emission 
associated with nebular shells.  
In some double-shell PNe, both shells exhibit a skin of enhanced 
{[O~{\sc iii}]/H$\alpha$}.  
The list of type B PNe includes NGC\,1501, NGC\,2371-2, NGC\,3242, 
NGC\,6153, NGC\,6818, NGC\,6826, NGC\,7009, and NGC\,7662 whose 
H$\alpha$ images and {[O~{\sc iii}]/H$\alpha$} ratio maps are shown 
in Fig.~\ref{fig_typeB}. 
Although some of these PNe have Fast Low-Ionization Emission Regions (the 
so-called FLIERs), the skins of enhanced {[O~{\sc iii}]/H$\alpha$} ratios 
described here are not related to these features.  
\item {\bf Type C} \\
PNe where the {[O~{\sc iii}]/H$\alpha$} ratio decreases at the outer edge 
of the nebula, i.e., these PNe show the opposite behavior to type B PNe.  
Among these PNe are: 
IC\,418, NGC\,6720, BD+30$^\circ$3639, MyCn\,18, Hen\,3-1357, NGC\,6881, 
NGC\,6886, M\,1-20, M\,2-14, M\,2-43, and Hen\,2-260.  
The H$\alpha$ image and {[O~{\sc iii}]/H$\alpha$} ratio map of these PNe 
are shown in Fig.~\ref{fig_typeC}. 
\item {\bf Type D} \\
PNe whose {[O~{\sc iii}]/H$\alpha$} ratio maps do not show 
the effects described above (Fig.~\ref{fig_typeD}).  
Some of them have flat {[O~{\sc iii}]/H$\alpha$} ratio maps (e.g., 
Wray\,16-286), some have low signal-to-noise (S/N) ratio maps 
(e.g., M\,3-14), and some of them show too complex structures (e.g., 
NGC\,2392 and NGC\,6369) to be clearly ascribed to type A, B, or C.  
\end{itemize}

The four different types described above are not mutually exclusive.  
In particular, most type A PNe are also type B, i.e., they show 
regions of enhanced {[O~{\sc iii}]/H$\alpha$} ratios that are 
associated both with expanding nebular shells and bow-shock features 
produced by fast collimated outflows.  
For these PNe, we use the type AB.  
The only exception is that of NGC\,6572, where a collimated outflow 
towards the north of the nebula produces a bow-shock feature with 
enhanced values of the {[O~{\sc iii}]/H$\alpha$} ratio, whereas this 
ratio is diminished around the main nebular shell.  
For this PN, we use the type AC.

The PNe with adequate [O~{\sc iii}]/H$\alpha$ ratio maps are listed 
in Table~\ref{tab2}.  
In this table we provide the classification of each PN and relevant 
information for a comparative study: 
the morphology or nebular shape (E=elliptical, B=bipolar, MP=quadrupolar 
or multi-polar, PS=point-symmetric, MS=multiple shell), 
the temperature of the central star, 
the nebular electron density, 
an averaged angular radius, the distance and linear radius, 
the [O~{\sc iii}] $\lambda$5007 and [N~{\sc ii}] $\lambda$6584 to H$\beta$ 
line intensity ratios, and comments on the presence of collimated outflows, 
FLIERs, and {\it ansae}, or on the quality of the observations.  
Most of the data in this table have been obtained from \citet{Frew2008}.  
Additional references are given in the last column of Table~\ref{tab2}.

\section{Discussion}

\subsection{Statistical Properties}

\begin{figure}[!t]
\begin{center}
\includegraphics[bb=18 100 592 720,width=1.0\columnwidth]{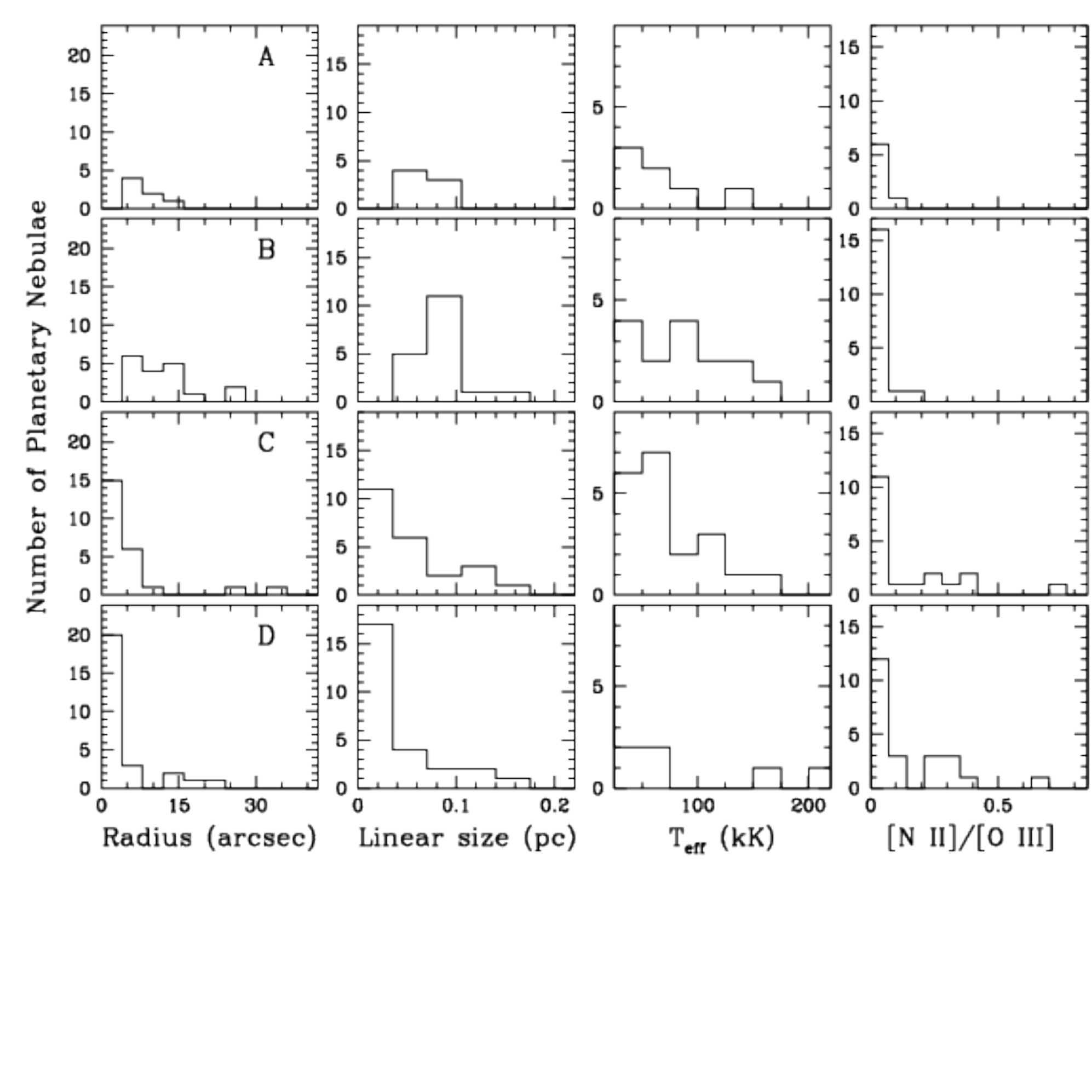}
\caption{
Distributions of the angular radii, linear sizes, 
{[N~{\sc ii}]/[O~{\sc iii}]} ratios, and stellar 
effective temperatures of the different types of PNe 
defined in Sect.~3.
}
\label{fig_histo}
\end{center}
\vspace*{-0.5cm}
\end{figure}

A preliminary inspection of Table~\ref{tab2} provides hints of the 
varying nature of PNe belonging to the different types described in 
the previous section.  
These are further illustrated by the distributions of different 
parameters shown in Fig.~\ref{fig_histo}.

Type D PNe is an heterogeneous group severely affected by 
several observational biases.
Most type D PNe are small, with angular radii $\leq$4\arcsec,  
and linear sizes $\leq$0.03 pc.  
The small size of these objects is clearly introducing an observational 
bias in the sample, as the fine morphological details of these objects 
cannot be properly resolved.  
Furthermore, the S/N of the images of a significant number of 
type D PNe is limited, thus hampering the detection of structures 
with contrasting values of the [O~{\sc iii}]/H$\alpha$ ratio.  
On the other hand, some objects in this group are large and well 
resolved, but they have very complex morphologies (e.g., NGC\,2392 
and NGC\,6369).  
Although these sources have low [N~{\sc ii}] $\lambda$6584 \AA\ to H$\alpha$ 
integrated line ratios, the presence of low-excitation features may result 
in local regions of significant [N~{\sc ii}] emission.  
The [N~{\sc ii}] contamination on the H$\alpha$ would correspond to 
features of apparently diminished values of the [O~{\sc iii}]/H$\alpha$ 
ratio.  
The complexity of the ratio maps may hide the possible effects of the 
expansion of nebular shells and collimated outflows in the surrounding 
material.

PNe of type C also tend to be small, with radii and linear sizes similar to, 
but slightly larger than, those of type D PNe.  
Notably, some of the type C PNe (e.g., NGC\,2346 and NGC\,6720) have 
the largest angular radii and linear sizes among the PNe in this 
sample.  
In contrast to type D PNe, the distribution of the effective 
temperature of the central stars of type C PNe clearly peaks 
at low values ($T_{\rm eff}\le50,000$~K).  
PNe of type C are either young, with low $T_{\rm eff}$ central stars 
(e.g., BD+30$^\circ$3639 and IC\,418), or have hot central stars 
which have already turned towards the white dwarf cooling track and 
reduced their stellar luminosities \citep[e.g., the evolved PN 
NGC\,6720 and the prominent bipolar PNe NGC\,2346 and 
NGC\,6886,][]{Napiwotzki99,MN81,PS05}.  
In the first case, the stellar spectrum is not hard enough to maintain the 
oxygen twice ionized throughout the whole nebula, whereas in the second 
case the spectrum is very hard but the photon flux seen by material at large 
distances from the central star is small.  
In both cases, the ionization degree drops at the edge of the nebula 
and an exterior O$^+$ region forms.  
Indeed, the reduced number of type C PNe with available \emph{HST} WFPC2 
images in the [O~{\sc ii}] $\lambda$3727 \AA\ line consistently reveal 
bright emission in regions of low {[O~{\sc iii}]/H$\alpha$}.  
The corresponding high value of the [O~{\sc ii}]/[O~{\sc iii}] line 
ratio in these regions can be interpreted as a sudden drop in the 
ionization degree.  
The decrease in the ionization degree will also enhance the emission 
of the low-excitation [N~{\sc ii}] lines.  
Therefore, the reduced {[O~{\sc iii}]/H$\alpha$} values observed in the ratio 
maps of IC\,418 \citep{R-L_etal2012} and NGC\,6720 \citep{ODell_etal2013} 
among others can be partially ascribed to the local contaminant contribution 
in these regions of [N~{\sc ii}] emission on the H$\alpha$ image.

PNe of type A and B are, on average, better resolved than type C PNe, with 
angular radii spanning between $\sim$4\arcsec\ and $\sim$20\arcsec, and 
linear sizes greater than 0.02 pc, peaking at $\sim$0.1 pc.  
The distribution of the effective temperature of their central 
stars (median value $\sim$75,000~K) is somehow similar to that 
of type C PNe (median value $\sim$60,000~K).  
More obvious differences are revealed by the distributions of their 
{[N~{\sc ii}]/[O~{\sc iii}]} ratios:  
those of type A and B peak at low values of the {[N~{\sc ii}]/[O~{\sc iii}]} 
ratio, with most data points below 0.1, whereas the distribution of type C PNe 
shows a notable tail towards high values of the {[N~{\sc ii}]/[O~{\sc iii}]} 
ratio.  
If we could exclude abundances effects, this result would imply 
that type C PNe have lower excitation than type A and B PNe.  
Finally, we remark that the occurrence of multiple-shell morphology and 
collimated outflows or FLIERs (suggestive of fast collimated outflows, 
although with expansion velocities comparable to these of the nebular 
shells) is the highest among PNe of type A and B.

\subsection{[O~{\sc iii}]/H$\alpha$ Radial Profiles}

\begin{figure}[!t]
\begin{center}
\includegraphics[bb=120 120 450 618,width=0.95\linewidth]{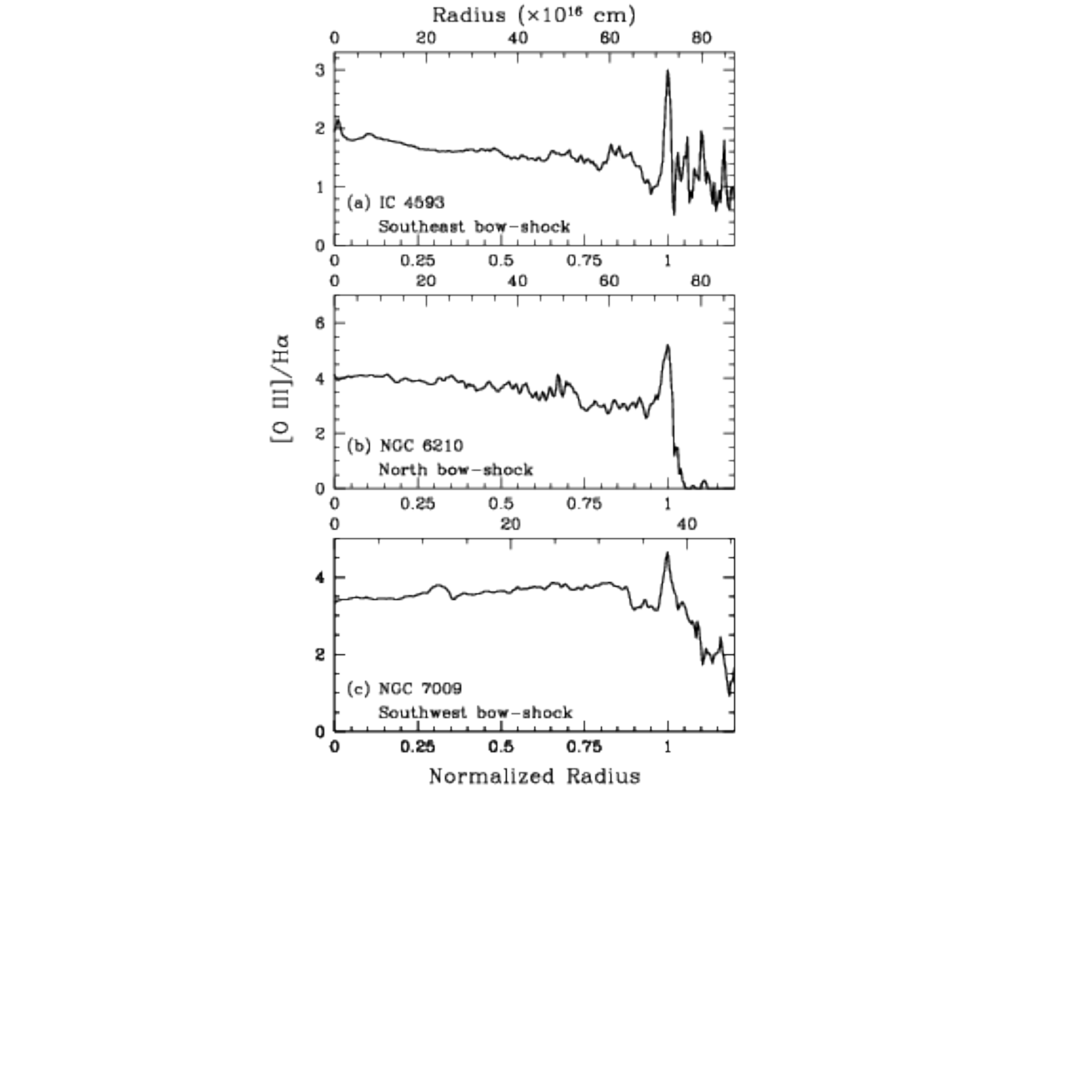}
\caption{
Radial profiles of the {[O~{\sc iii}]/H$\alpha$} ratio maps of the type 
A PNe IC\,4593, NGC\,6210, and NGC\,7009 along selected bow-shock features 
associated with collimated outflows.  
}
\label{profA}
\end{center}
\end{figure}

\begin{figure}[!t]
\begin{center}
\includegraphics[bb=120 30 450 618,width=0.90\linewidth]{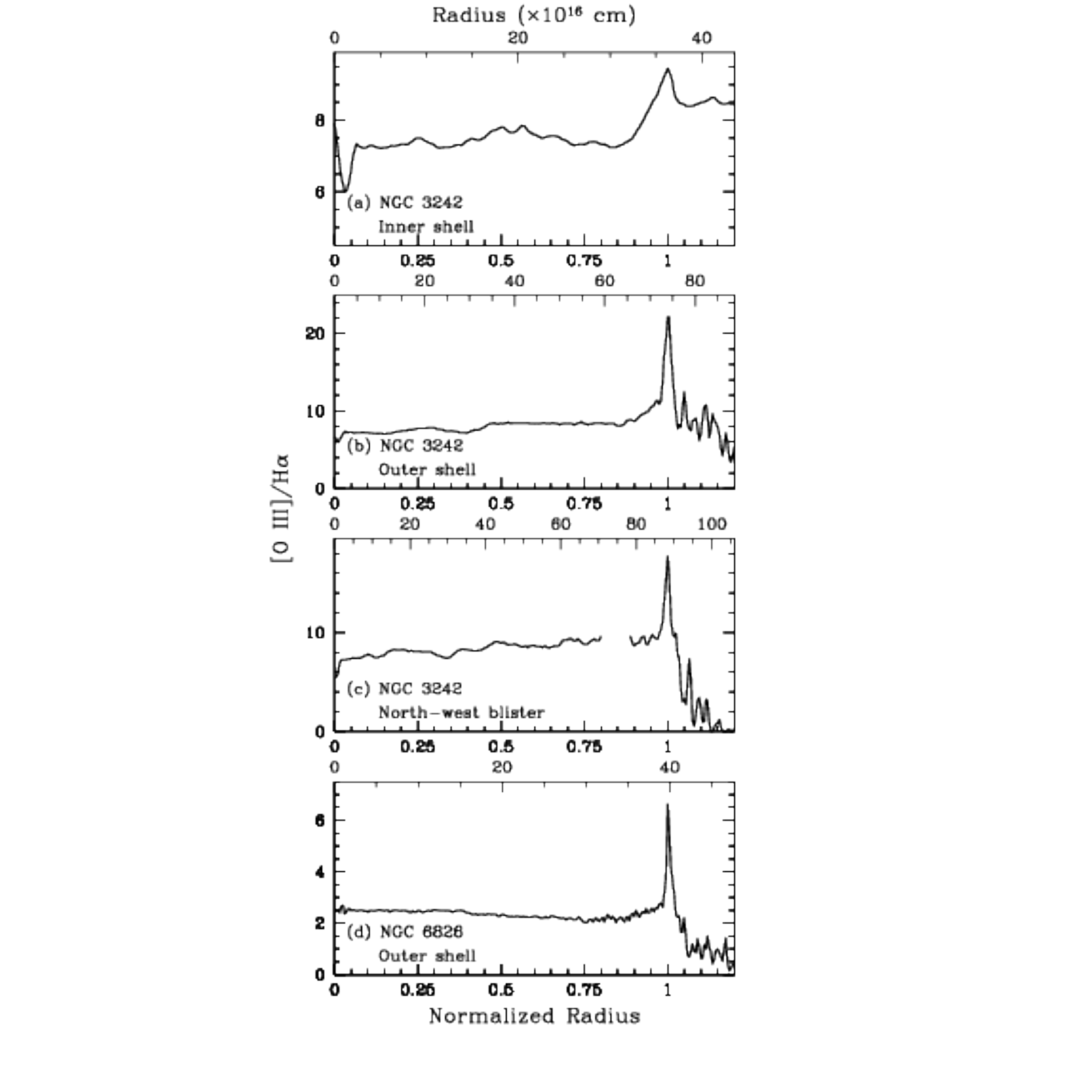}
\caption{
Radial profiles of the {[O~{\sc iii}]/H$\alpha$} ratio maps of the type 
B PNe NGC\,3242 {\it (a,b,c)} and NGC\,6826 {\it (d)}.  
The radial profiles of NGC\,3242 correspond to the inner shell {\it (a)}, 
the outer shell {\it (b)}, and the north-west blister of the outer shell 
{\it (c)}.  
The radial profile of NGC\,6826 {\it (d)} corresponds to the outer shell.  
The [O~{\sc iii}]/H$\alpha$ peaks of the outer shell of NGC\,3242 
are not shown in the profile of the blister of this shell {\it (c)}.  
}
\label{profB}
\end{center}
\end{figure}

\begin{figure}[!t]
\begin{center}
\includegraphics[bb=120 120 450 618,width=0.95\linewidth]{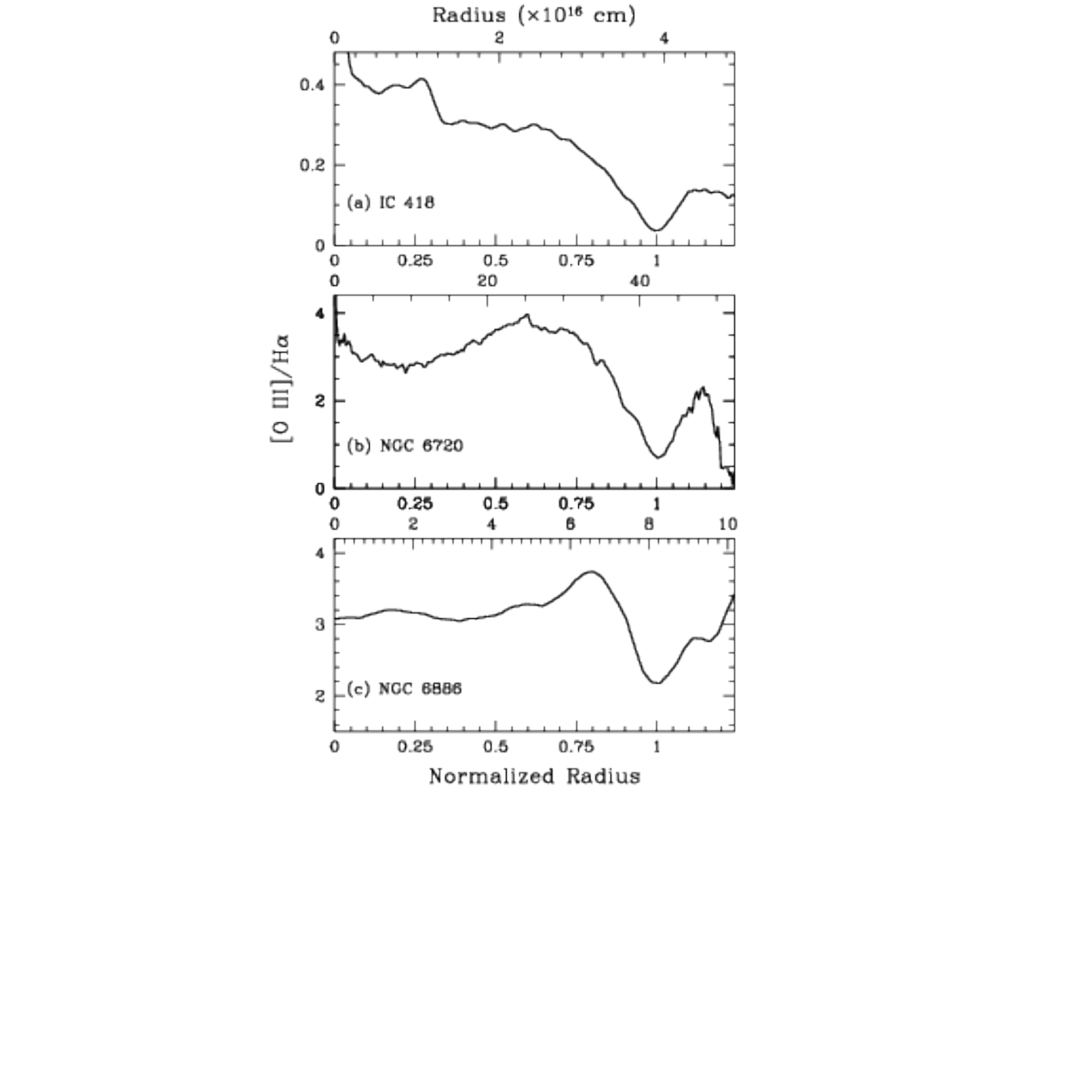}
\caption{
Radial profiles of the {[O~{\sc iii}]/H$\alpha$} ratio maps of the 
type C PNe IC\,418, NGC\,6720, and NGC\,6886 along their minor axes 
at PA$\sim$70$^\circ$, 130$^\circ$, and 45$^\circ$, respectively.  
}
\label{profC}
\end{center}
\end{figure}

To investigate in more detail the physical properties of the regions where 
the variations of the {[O~{\sc iii}]/H$\alpha$} ratio take place, we have 
extracted radial profiles of representative PNe of types A, B, and C.  
These radial profiles, shown in Figs.~\ref{profA}, \ref{profB}, and 
\ref{profC}, respectively, have been derived from flux calibrated 
\emph{HST} images;  however, we have not attempted to deredden the 
line ratios and, therefore, the values of the {[O~{\sc iii}]/H$\alpha$} 
ratio in these profiles do not correspond to absolute line intensity 
ratios.  
The relative enhancement (or decrement for type C PNe) of the 
{[O~{\sc iii}]/H$\alpha$} ratio and the width of the region with 
anomalous [O~{\sc iii}] emission derived from these profiles 
are listed in Table~\ref{tab3}.

The regions of enhanced {[O~{\sc iii}]/H$\alpha$} ratio associated with 
bow-shocks, outer shells, and blisters show the largest increase of this 
ratio, being typically larger by a factor of two.  
On the contrary, the enhancement of the [O~{\sc iii}]/H$\alpha$ 
ratio at the inner shells of type B PNe is typically smaller.  
Those inner shells are embedded within the corresponding outer shells, and thus 
we interpret that the smaller enhancement of the [O~{\sc iii}]/H$\alpha$ ratio 
is the result of the dilution of the emission from the skin of high 
[O~{\sc iii}]/H$\alpha$ ratio by the emission from the outer shell.

There are also notable differences among the fractional sizes of these 
regions.  
The peaks of enhanced [O~{\sc iii}] emission in bow-shocks, 
outer shells, and blisters show the narrowest fractional 
sizes, all being $\le$4\%.  
Meanwhile, the regions of enhanced {[O~{\sc iii}]/H$\alpha$} ratios 
associated with the inner shell of NGC\,3242 and the absorptions in the 
{[O~{\sc iii}]/H$\alpha$} radial profiles of type C PNe are broader, 
with fractional widths above 10\%.  
The physical sizes of all these regions are rather similar, 
$\approx$10$^{16}$ cm, but for the outer 
shell of NGC\,6826, which is especially thin (4$\times$10$^{15}$ 
cm).  
Even the outer regions of diminished {[O~{\sc iii}]/H$\alpha$} ratios 
of type C have similar physical sizes, except for NGC\,6720, which is 
especially broad ($\approx$10$^{17}$ cm).

There are two interesting features in these radial profiles which are worth 
mentioning as they are indicative of the variations in excitation affecting 
the nebular material and of the geometry of these regions.  
First, the radial profiles of bow-shocks in Fig.~\ref{profA} 
display a depression just inside the emission peak which is then 
followed by a bounce in this ratio further in.  
The regions of diminished values in the {[O~{\sc iii}]/H$\alpha$} ratio 
are associated with the prominent low-ionization features found inside 
these bow-shocks, whereas the increase in the value of the line ratio 
suggests a relative brightening of the [O~{\sc iii}] emission on larger 
spatial scales than the outer skin. 
Secondly, the {[O~{\sc iii}]/H$\alpha$} peak associated with outer shells 
is preceded by a ramp in this ratio, as can be seen in the radial profiles 
of the outer shells of NGC\,3242 and NGC\,6826 (Fig.~\ref{profB}).  
Projection effects can be responsible for this feature.

\subsection{Origin of the High [O~{\sc iii}]/H$\alpha$ Skins}

As mentioned in Sect.\ 4.1, the [O~{\sc iii}]/H$\alpha$ ratio maps of type 
C PNe can be explained in terms of a nebular ionization structure in which 
the O$^{++}$/H$^+$ ratio decreases in the outermost nebular layers.  
The low effective temperature of the central stars of young PNe or 
the low luminosity of the central stars of bipolar or evolved PNe 
may result in these low ionization degrees.

\begin{figure}[!t]
\begin{center}
\includegraphics[width=0.99\linewidth]{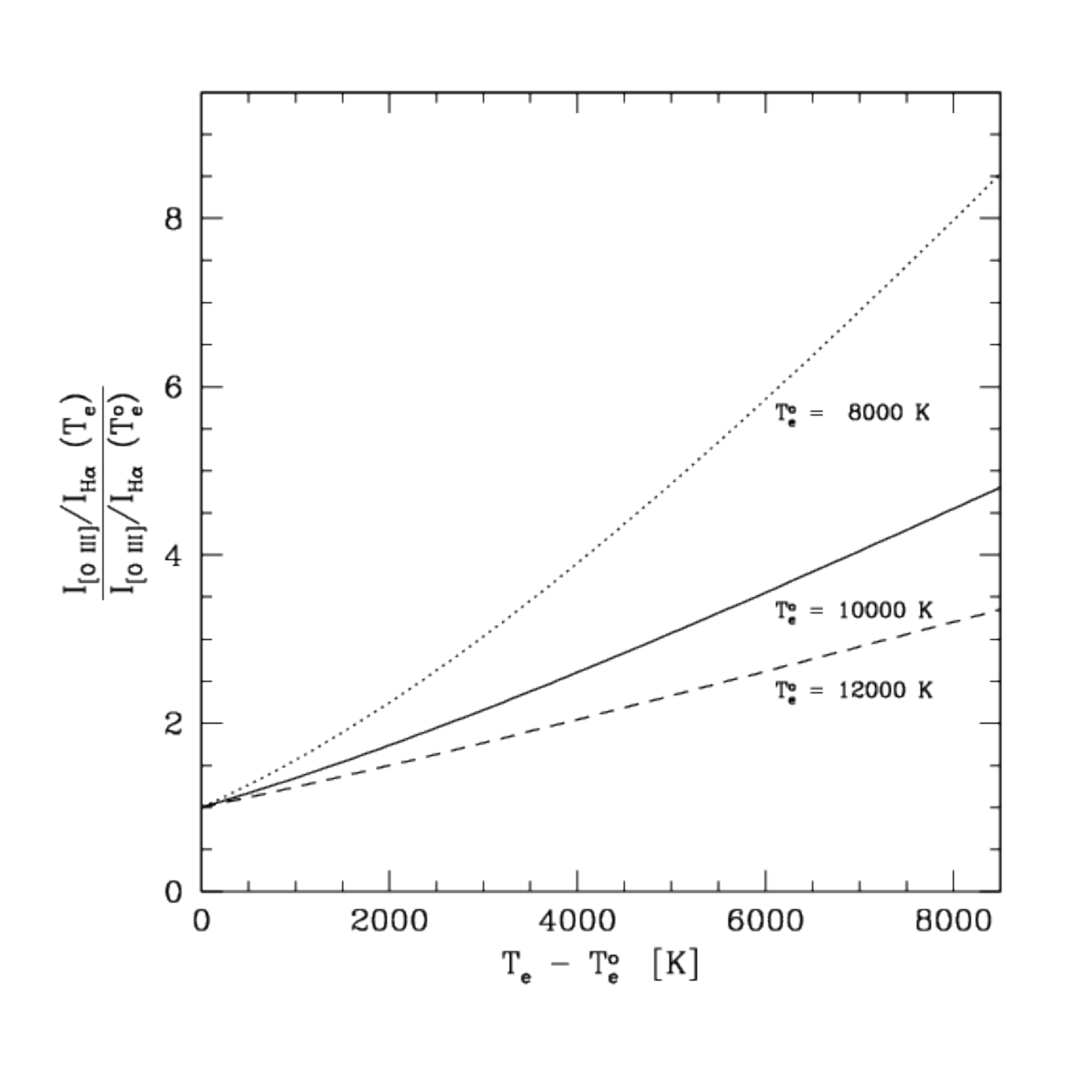}
\caption{
Variation of the {[O~{\sc iii}]/H$\alpha$} line intensity ratio 
associated with the increase in electron temperature from $T_{\rm e}^{\rm o}$ 
to $T_{\rm e}$.  
Three different curves are plotted, depending on the value of the 
pre-shock temperature.  
}
\label{te_plot}
\end{center}
\end{figure}

Similarly, one might seek an explanation to the [O~{\sc iii}]/H$\alpha$ 
ratio maps of type A and B PNe in terms of a nebular ionization structure: 
the outward increase in the [O~{\sc iii}] emission would be caused by 
a decrease of the O$^{3+}$/H$^+$ ratio and a subsequent increase in the 
O$^{++}$/H$^+$ ratio. 
If this were the case, one would expect low [O~{\sc iii}] emission 
throughout the nebula and its brightening at the outer edge, 
mimicking the structure of the [O~{\sc ii}] emission of type C PNe.  
However, it turns out that the brightness of the [O~{\sc iii}] $\lambda$5007 
\AA\ line is consistently high throughout the whole nebula in both type A 
and B PNe. 
Furthermore, the O$^{3+}$/O$^{++}$ fraction required to reproduce such a 
structure would require stellar temperature and luminosities much larger 
than those typical of PNe (the maximum O$^{3+}$/O$^{++}$ values observed 
in PNe are about 0.5). 
This indicates that the local enhancement of the [O~{\sc iii}]/H$\alpha$ 
ratio in type A and B PNe is most probably caused by a sharp increase 
in temperature, which cannot be reproduced by photoionization models, 
unless an \emph{ad hoc} additional heating mechanism working on small 
spatial scales produces a sharp raise in $T_{\rm e}$.  
Such an increase in $T_{\rm e}$ would reduce the effective recombination 
rate of H$\alpha$, which is roughly proportional to $T_{\rm e}^{-0.9}$, 
and increase the collisional excitation of O$^{++}$, which is roughly 
proportional to $T^{-1/2}\,e^{-E/kT}$.  
The [O~{\sc iii}]/H$\alpha$ line intensity ratio can be expressed as: \\
\begin{equation}
\hspace*{0.5cm}
I_{[\mathrm{O~III}]}/I_{\mathrm{H}\alpha} \propto \frac{N(\mathrm{O}^{++})}{N(\mathrm{H}^+)} \; T_e^{0.4} \; e^{-28,000/T_e}
\end{equation}
We plot in Fig.~\ref{te_plot} the variation of the [O~{\sc iii}]/H$\alpha$ 
line intensity ratio with the increase of electron temperature above a 
baseline temperature in the range typical of PNe, 8,000~K$\leq T_e 
\leq$12,000~K.  
We note that these curves have not been computed using the 
above expression, but the exact line emissivities.  
For a baseline temperature of the nebular material of 10,000~K, the 
observed increase of the [O~{\sc iii}]/H$\alpha$ line intensity ratio 
by factors of 2 to 3 implies a jump in $T_{\rm e}$ of 2,000--5,000~K.

The extra heating can be provided by a forward shock propagating into 
a tenuous medium \citep{Cox1972} as it has been observed in collimated 
outflows in Herbig-Haro objects \citep[e.g., HH\,34, ][]{R-Getal12} 
and ring nebulae around WR stars \citep{Gruendl_etal2000}. 
The extra heating in type A PNe is provided by the forward shock 
of a fast outflow propagating into the tenuous, outer regions of 
the PNe or the interstellar medium (ISM).  
The shocks produced by collimated outflows in Herbig-Haro objects enhance 
the emission of low-ionization lines (e.g., [O~{\sc i}], [S~{\sc ii}], or 
[N~{\sc ii}]), as predicted for steady, plane-parallel shock models 
\citep{HRH87}, but the situation is notably different in PNe.  
Their outflows are photoionized by the central star, and this results 
in variations of the expected line ratios and spatial distribution of 
the emission from different lines that has been modeled for fast, 
compact knots traveling away from the central star by \citet{Raga_etal08}. 
In particular, these models predict a relative brightening of the 
[O~{\sc iii}] emission with respect to that of H$\alpha$ at larger 
distances from the ionizing source (their Fig.~9), in clear 
agreement with the radial profiles displayed in Fig.~\ref{profA}.

Meanwhile, the expansion of a higher density shell into a lower 
density, outer shell or into the ISM produces a forward shock 
in type B PNe which heats the material in front of the shell.  
In this case, the pre-shock material is already highly ionized by the central 
star of the PN, and no enhancement of the emission from low-ionization lines 
is expected.  
Therefore, the analysis of [O~{\sc iii}]/H$\alpha$ ratio maps 
provides the only strategy to investigate the occurrence of 
the shocks associated to expanding shells of PNe.

%

Hydrodynamical models of PN evolution indeed predict the development of two 
different shocks during their formation \citep[e.g., ][]{Petal04,Setal05}.  
The sudden ionization of the material ejected during the AGB 
produces a D-type ionization front at the leading edge of the 
outer nebular shell of multiple-shell PNe.  
This shock propagates outwards, into the ambient AGB wind, and a 
thin skin of shocked material is expected at the leading shock of 
the shell.  
Model simulations of the propagation of such shock kindly performed by 
D.\ Sch\"onberner and M.\ Steffen suggest that the post-shock temperatures 
depend very sensitively on the density jump across the shock, although a 
high numerical resolution is required to resolve the shock.  
Similarly, another shock is set by the wind interaction at the discontinuity 
between the outer nebular shell and the bright rim pushed by the thermal 
pressure of the hot bubble.  
The available model simulations predict the formation of a thick rim 
early in the evolution of the PN, resulting in a shock too weak to 
produce the observed enhancement of the [O~{\sc iii}]/H$\alpha$ ratios.  
With time, the rim sharpens and moves faster, and its leading shock 
becomes stronger \citep{Petal04}.

These detailed hydrodynamical simulations determine that the 
velocity jump across the shock front at the leading edge of 
the outer shell is 5--40~km~s$^{-1}$.  
Across the shock front between the inner rim and the shell, the velocity 
jump is smaller, $\leq10$~km~s$^{-1}$.  
For typical electron densities of $\sim$4,000 cm$^{-3}$, $\sim$1,000 cm$^{-3}$, 
and $\sim$50 cm$^{-3}$ for the rim, outer shell, and surrounding unperturbed 
AGB wind, these velocity jumps imply temperature increases $\simeq$4,000~K 
which can easily produce the observed enhancement of the 
[O~{\sc iii}]/H$\alpha$ ratios.



\section{Summary}

Using archival \emph{HST} [O~{\sc iii}] and H$\alpha$ images of PNe, we have 
found regions of enhanced values of the [O~{\sc iii}]/H$\alpha$ ratio.  
These regions can be described as thin skins which envelop either 
bow-shock features associated with fast collimated outflows PNe 
or nebular shells of PNe with multiple-shell morphology.  
The enhancement of the [O~{\sc iii}]/H$\alpha$ ratio is interpreted to be 
caused by a local increase of $T_\mathrm{e}$ which brightens the emission 
in the [O~{\sc iii}] line and damps the emissivity of the H$\alpha$ line.  
The amplitude of the variations in the values of the [O~{\sc iii}]/H$\alpha$ 
ratio implies variations in $T_\mathrm{e}$ of 2,000--5,000~K.  
The increase in $T_\mathrm{e}$ is associated with forward shocks 
produced by the expansion of collimated outflows and nebular 
shells into surrounding media of lower density.  
These results illustrate the complexity of PNe and its suitability 
to studying the detailed physics of ionized plasmas.

\begin{acknowledgements}
All of the data presented in this paper were obtained from the Mikulski 
Archive for Space Telescopes (MAST). 
STScI is operated by the Association of Universities for Research in 
Astronomy, Inc., under NASA contract NAS5-26555. 
Support for MAST for non-HST data is provided by the NASA Office of 
Space Science via grant NNX09AF08G and by other grants and contracts.
This work is funded by grants AYA 2005-01495 of the Spanish MEC 
(Ministerio de Educaci\'on y Ciencia), grant FGM-1747 of the Junta 
de Andaluc\'\i a, and CSIC grant 2006-50-I135.
MAG and JAT are supported by the Spanish MICINN (Ministerio de Ciencia e 
Innovaci\'on) grant AYA 2011-29754-C03-02, VL by grant AYA 2011-22614, 
LFM by grant AYA 2011-30228-C03-01, and AR by grant AYA 2011-30228-C03, 
all of them co-funded with FEDER funds.  
JAT acknowledges support by the CSIC JAE-Pre student grant 2011-00189, 
and PVF by CONACyT grant 167611 and DGAPA-PAPIIT (UNAM) grant IN105312.  
We kindly thanks Detlef Sch\"onberner and Matthias Steffen for helpful 
discussion.  
\end{acknowledgements}

\begin{table*}
\centering
\scriptsize{
\caption{\footnotesize{
Sample of PNe with \emph{HST} WFPC2 or WFC3 H$\alpha$ and [O~{\sc iii}] 
observations}}

\label{tab1}
\begin{tabular}{llrrr|llrrr}
\hline\hline

\multicolumn{1}{l}{PN G} & 
\multicolumn{1}{l}{Common Name} & 
\multicolumn{1}{c}{H$\alpha$} &  
\multicolumn{1}{c}{[O~{\sc iii}]} &  
\multicolumn{1}{r}{Proposal ID} & 
\multicolumn{1}{l}{PN G} & 
\multicolumn{1}{l}{Common Name} & 
\multicolumn{1}{c}{H$\alpha$} &  
\multicolumn{1}{c}{[O~{\sc iii}]} &  
\multicolumn{1}{r}{Proposal ID} \\
\multicolumn{1}{l}{} & 
\multicolumn{1}{l}{} & 
\multicolumn{1}{c}{Exp.\ Time} &  
\multicolumn{1}{c}{Exp.\ Time} &  
\multicolumn{1}{l}{} & 
\multicolumn{1}{l}{} & 
\multicolumn{1}{l}{} & 
\multicolumn{1}{c}{Exp.\ Time} &  
\multicolumn{1}{c}{Exp.\ Time} &  
\multicolumn{1}{l}{} \\  
\multicolumn{1}{l}{} & 
\multicolumn{1}{l}{} & 
\multicolumn{1}{c}{(s)} & 
\multicolumn{1}{c}{(s)} & 
\multicolumn{1}{l}{} & 
\multicolumn{1}{l}{} & 
\multicolumn{1}{l}{} & 
\multicolumn{1}{c}{(s)} & 
\multicolumn{1}{c}{(s)} & 
\multicolumn{1}{l}{} \\

\hline

000.3$+$12.2 & IC\,4634                         & 1000~~~~~~ & 1000~~~~~~ & 6856             & 084.2$+$01.0 & K\,4-55                          & 2460~~~~~~ & 2440~~~~~~ & 11956            \\ 
001.2$+$02.1 & Hen\,2-262                       &  280~~~~~~ &  280~~~~~~ & 9356             & 084.9$-$03.4 & NGC\,7027\tablefootmark{c}       &  500~~~~~~ &  100~~~~~~ & 11122            \\     
001.7$-$04.4 & H\,1-55                          &  200~~~~~~ &  280~~~~~~ & 9356             & 089.8$-$05.1 & IC\,5117                         &  240~~~~~~ &  320~~~~~~ & 8307             \\     
002.3$-$03.4 & H\,2-37                          &  280~~~~~~ &  280~~~~~~ & 9356             & 096.4$+$29.9 & NGC\,6543                        &  800~~~~~~ & 1600~~~~~~ & 5403             \\     
002.4$+$05.8 & NGC\,6369                        &  640~~~~~~ &  640~~~~~~ & 9582             & 106.5$-$17.6 & NGC\,7662                        &  200~~~~~~ &  500~~~~~~ & 6117, 6943, 8390 \\     
002.7$-$04.8 & M\,1-42                          &  900~~~~~~ & 1800~~~~~~ & 11185            & 111.8$-$02.8 & Hb\,12                           & 1600~~~~~~ & 1600~~~~~~ & 11093            \\  
002.9$-$03.9 & H\,2-39                          &  280~~~~~~ &  280~~~~~~ & 9356             & 138.8$+$02.8 & IC\,289                          & 2000~~~~~~ & 2000~~~~~~ & 11956            \\    
003.5$-$04.6 & NGC\,6565                        &  160~~~~~~ &  320~~~~~~ & 11122            & 144.1$+$06.1 & NGC\,1501                        & 1600~~~~~~ & 2000~~~~~~ & 11956            \\  
003.6$+$03.1 & M\,2-14                          &  280~~~~~~ &  280~~~~~~ & 9356             & 189.1$+$19.8 & NGC\,2371-72                     & 1600~~~~~~ & 1600~~~~~~ & 11093            \\     
003.8$+$05.3 & H\,2-15                          &  280~~~~~~ &  280~~~~~~ & 9356             & 197.8$+$17.3 & NGC\,2392                        &  400~~~~~~ &  400~~~~~~ & 8499             \\     
003.9$-$03.1 & KFL\,7                           &  280~~~~~~ &  280~~~~~~ & 9356             & 215.2$-$24.2 & IC\,418                          &  888~~~~~~ &  360~~~~~~ & 6353, 7501       \\     
004.0$-$03.0 & M\,2-29                          &  200~~~~~~ &  160~~~~~~ & 9356             & 231.8$+$04.1 & NGC\,2438                        & 2080~~~~~~ & 2080~~~~~~ & 11827            \\     
004.1$-$03.8 & KFL\,11                          &  280~~~~~~ &  280~~~~~~ & 9356             & 215.6$+$03.6 & NGC\,2346                        &  200~~~~~~ &  120~~~~~~ & 7129             \\     
004.8$-$22.7 & Hen\,2-436                       &  200~~~~~~ &  160~~~~~~ & 9356             & 234.8$+$02.4 & NGC\,2440                        & 1600~~~~~~ & 1600~~~~~~ & 11090            \\     
004.8$+$02.0 & H\,2-25                          &  400~~~~~~ &  400~~~~~~ & 9356             & 249.0$+$06.9 & SaSt\,1-1                        &  200~~~~~~ &  280~~~~~~ & 8332             \\     
005.2$-$18.6 & StWr\,2-21                       &  280~~~~~~ &  280~~~~~~ & 9356             & 261.0$+$32.0 & NGC\,3242                        &  100~~~~~~ & 1200~~~~~~ & 6117, 7501, 8773 \\     
006.1$+$08.3 & M\,1-20                          &  200~~~~~~ &  160~~~~~~ & 9356             & 261.9$+$08.5 & NGC\,2818                        & 1600~~~~~~ & 2000~~~~~~ & 11956            \\     
006.3$+$04.4 & H\,2-18                          &  280~~~~~~ &  280~~~~~~ & 9356             & 272.1$+$12.3 & NGC\,3132                        &  400~~~~~~ & 1200~~~~~~ & 6221, 8390       \\     
006.4$+$02.0 & M\,1-31                          &  780~~~~~~ &  160~~~~~~ & 9356             & 285.6$-$02.7 & Hen\,2-47                        & 1600~~~~~~ & 1600~~~~~~ & 11090            \\     
006.8$-$19.8 & Wray 16-423                      &  200~~~~~~ &  160~~~~~~ & 9356             & 285.7$-$14.9 & IC\,2448                         &  200~~~~~~ &  320~~~~~~ & 11122            \\     
006.8$+$04.1 & M\,3-15                          &  200~~~~~~ &  160~~~~~~ & 9356             & 294.6$+$04.7 & NGC\,3918                        &  140~~~~~~ &  320~~~~~~ & 11122            \\     
007.5$+$04.3 & Th\,4-1                          &  280~~~~~~ &  280~~~~~~ & 9356             & 305.1$+$01.4 & Hen\,2-90                        & 2325~~~~~~ & 1210~~~~~~ & 8345, 9102       \\     
008.2$+$06.8 & Hen\,2-260                       &  200~~~~~~ &  460~~~~~~ & 9356             & 307.5$-$04.9 & MyCn\,18                         &  600~~~~~~ & 1400~~~~~~ & 6221             \\     
008.6$-$02.6 & MaC 1-11                         &  280~~~~~~ &  280~~~~~~ & 9356             & 309.1$-$04.3 & NGC\,5315                        & 1600~~~~~~ & 1600~~~~~~ & 11090            \\     
009.3$+$05.7 & Hen\,3-1475                      &  830~~~~~~ &  800~~~~~~ & 7285             & 312.3$+$10.5 & NGC\,5307                        & 1600~~~~~~ & 1600~~~~~~ & 11090            \\     
010.0$+$00.7 & NGC\,6537                        & 1240~~~~~~ & 1000~~~~~~ & 6502             & 319.6$+$15.7 & IC\,4406\tablefootmark{d}        &  540~~~~~~ &  600~~~~~~ & 8726, 9314       \\     
010.8$+$18.0 & M\,2-9                           & 1240~~~~~~ & 1000~~~~~~ & 6502             & 324.0$+$03.5 & PM\,1-89                         & 4900~~~~~~ & 2900~~~~~~ & 5404, 5864       \\     
010.8$-$01.8 & NGC\,6578                        &  160~~~~~~ &  320~~~~~~ & 11122            & 327.8$+$10.8 & NGC\,5882                        &  140~~~~~~ &  380~~~~~~ & 11122            \\  
019.4$-$05.3 & M\,1-61                          &  240~~~~~~ &  320~~~~~~ & 8307             & 331.1$-$05.7 & PC\,11                           &  200~~~~~~ &  280~~~~~~ & 8332             \\   
025.3$+$40.8 & IC\,4593                         & 1600~~~~~~ & 1600~~~~~~ & 11093            & 331.3$-$12.1 & Hen\,3-1357                      &  240~~~~~~ &  368~~~~~~ & 6039, 8390       \\    
025.8$-$17.9 & NGC\,6818                        &  520~~~~~~ & 1300~~~~~~ & 6792, 7501, 8773 & 331.7$-$01.0 & Mz\,3\tablefootmark{e}           & 1260~~~~~~ & 1160~~~~~~ & 6856, 9050       \\     
027.6$+$04.2 & M\,2-43                          &  520~~~~~~ & 1800~~~~~~ & 8307             & 341.8$+$05.4 & NGC\,6153                        & 1000~~~~~~ & 1200~~~~~~ & 8594             \\     
034.6$+$11.8 & NGC\,6572                        &  180~~~~~~ &  840~~~~~~ & 7501, 9839       & 349.5$+$01.0 & NGC\,6302\tablefootmark{a}       & 2100~~~~~~ & 2220~~~~~~ & 11504            \\     
036.1$-$57.1 & NGC\,7293                        & 1800~~~~~~ & 1800~~~~~~ & 5977             & 351.1$+$04.8 & M\,1-19                          &  160~~~~~~ &  160~~~~~~ & 9356             \\     
037.7$-$34.5 & NGC\,7009                        &  400~~~~~~ &  320~~~~~~ & 8114             & 351.9$-$01.9 & Wray\,16-286                     &  200~~~~~~ &  280~~~~~~ & 9356             \\     
037.8$-$06.3 & NGC\,6790                        &  160~~~~~~ &  200~~~~~~ & 8307             & 352.6$+$03.0 & H\,1-8                           &  200~~~~~~ &  280~~~~~~ & 9356             \\     
043.1$+$37.7 & NGC\,6210                        &  320~~~~~~ &  320~~~~~~ & 6792             & 353.5$-$05.0 & JaFu\,2\tablefootmark{f}         & 3600~~~~~~ & 2000~~~~~~ & 6780             \\     
054.1$-$12.1 & NGC\,6891                        & 1280~~~~~~ &  320~~~~~~ & 11122            & 354.5$+$03.3 & Th\,3-4                          &  280~~~~~~ &  280~~~~~~ & 9356             \\  
054.2$-$03.4 & Necklace Nebula\tablefootmark{a} & 2000~~~~~~ & 2000~~~~~~ & 12675            & 354.9$+$03.5 & Th\,3-6                          &  280~~~~~~ &  400~~~~~~ & 9356             \\  
057.9$-$01.5 & Hen\,2-447                       &  520~~~~~~ & 1800~~~~~~ & 8307             & 355.4$-$02.4 & M\,3-14                          &  200~~~~~~ &  160~~~~~~ & 9356             \\    
060.1$-$07.7 & NGC\,6886                        & 1120~~~~~~ & 1020~~~~~~ & 7501, 8345, 8773 & 355.9$+$03.6 & H\,1-9                           &  280~~~~~~ &  280~~~~~~ & 9356             \\     
060.8$-$03.6 & NGC\,6853                        & 2000~~~~~~ & 1000~~~~~~ & 8726             & 356.1$-$03.3 & H\,2-26                          &  280~~~~~~ &  280~~~~~~ & 9356             \\         
063.1$+$13.9 & NGC\,6720                        &  480~~~~~~ &  720~~~~~~ & 7632, 8726       & 356.5$-$03.6 & H\,2-27                          &  360~~~~~~ &  400~~~~~~ & 9356             \\     
064.1$+$04.3 & M\,1-92                          &  680~~~~~~ & 2080~~~~~~ & 6533             & 356.9$+$04.4 & M\,3-38                          &  280~~~~~~ &  280~~~~~~ & 9356             \\     
064.7$+$05.0 & BD+30$^\circ$3639                 &  484~~~~~~ &  900~~~~~~ & 8116, 8390       &357.1$-$04.7 & H\,1-43                          &  200~~~~~~ &  280~~~~~~ & 9356             \\     
065.0$-$27.3 & Ps\,1\tablefootmark{b}           &11420~~~~~~ & 1040~~~~~~ & 6751             & 357.2$+$02.0 & H\,2-13                          &  280~~~~~~ &  280~~~~~~ & 9356             \\     
071.6$-$02.3 & M\,3-35                          &  520~~~~~~ & 1000~~~~~~ & 8307             & 358.5$-$04.2 & H\,1-46                          &  160~~~~~~ &  160~~~~~~ & 9356             \\     
073.0$-$02.4 & K\,3-76                          &    6~~~~~~ &   18~~~~~~ & 6943             & 358.5$+$02.9 & Wray\,16-282                     &  280~~~~~~ &  280~~~~~~ & 9356             \\  
074.5$+$02.1 & NGC\,6881                        &  280~~~~~~ &  320~~~~~~ & 8307             & 358.9$+$03.4 & H\,1-19                          &  200~~~~~~ &  280~~~~~~ & 9356             \\     
082.1$+$07.0 & NGC\,6884                        & 1100~~~~~~ &  560~~~~~~ & 8345, 8390       & 359.2$+$04.7 & Th\,3-14                         &  280~~~~~~ &  400~~~~~~ & 9356             \\     
082.5$+$11.3 & NGC\,6833                        &   40~~~~~~ &    3~~~~~~ & 6943, 6353       & 359.3$-$00.9 & Hb\,5                            & 1300~~~~~~ & 1000~~~~~~ & 6502             \\     
083.5$+$12.7 & NGC\,6826                        &  100~~~~~~ &  100~~~~~~ & 6117             &                                                 &            &            &                  \\

\hline                                          
\end{tabular}                                   
\tablefoot{\\
\tablefoottext{a}{
Source observed with the WFC3 camera. \\ }
\tablefoottext{b}{
PN in the globular cluster NGC\,7078. \\ }
\tablefoottext{c}{
The nebula is mostly covered by the WFPC2-PC1 chip, but no completely. \\ }
\tablefoottext{d}{
The nebula is covered by two adjacent WFPC2 observations. \\ }
\tablefoottext{e}{
Possible symbiotic star. \\ }
\tablefoottext{f}{
PN in the globular cluster NGC\,6441. } 
}}
\end{table*}

\begin{table*}[!t]
\centering
\scriptsize{
\caption{\footnotesize{[O~{\sc iii}]/H$\alpha$ Type and Basic Parameters of Planetary Nebulae} 
\vspace*{-0.3cm}}
\label{tab2}
\begin{tabular}{lllrrrrrrrll}
\hline\hline
\multicolumn{1}{l}{Common} & 
\multicolumn{1}{c}{[O~{\sc iii}]/H$\alpha$} &  
\multicolumn{1}{c}{Shape} & 
\multicolumn{1}{c}{$T_{\rm eff}$} & 
\multicolumn{1}{c}{$N_{\rm e}$} & 
\multicolumn{1}{c}{$r$} & 
\multicolumn{1}{c}{$d$} & 
\multicolumn{1}{c}{$R$} & 
\multicolumn{1}{c}{[O~{\sc iii}]/H$\beta$} & 
\multicolumn{1}{c}{[N~{\sc ii}]/H$\beta$} & 
\multicolumn{1}{l}{Comments} & 
\multicolumn{1}{l}{References} \\
\multicolumn{1}{l}{Name} & 
\multicolumn{1}{c}{Class} & 
\multicolumn{1}{c}{} & 
\multicolumn{1}{c}{} & 
\multicolumn{1}{c}{} & 
\multicolumn{1}{c}{} & 
\multicolumn{1}{c}{} & 
\multicolumn{1}{c}{} & 
\multicolumn{1}{c}{} & 
\multicolumn{1}{c}{} & 
\multicolumn{1}{c}{} & 
\multicolumn{1}{c}{} \\
\multicolumn{1}{l}{} & 
\multicolumn{1}{c}{} & 
\multicolumn{1}{c}{} & 
\multicolumn{1}{c}{(kK)} & 
\multicolumn{1}{c}{(cm$^{-3}$)} & 
\multicolumn{1}{c}{($\arcsec$)} & 
\multicolumn{1}{c}{(kpc)} & 
\multicolumn{1}{c}{(pc)} & 
\multicolumn{1}{c}{} & 
\multicolumn{1}{c}{} & 
\multicolumn{1}{c}{} & 
\multicolumn{1}{c}{} \\
\hline
IC\,4593          & AB & MS     &     40~~ & 3,000 &  7.5 &   1.57 & 0.06  &   560 &    12 & PS outflows & 1,2  \\ 
IC\,4634          & A  & E, MS  &     49~~ & 2,700 &  4.2 &   2.8  & 0.06  &   999 &    11 & PS outflows & 3,4  \\
NGC\,3918         & AB & E, MS  &    150~~ & 5,500 &  9.0 &   1.84 & 0.08  &  1657 &    84 & PS outflows & 1,5  \\
NGC6\,6210        & A  & PS, MS &     60~~ & 5,100 &  8.0 &   2.10 & 0.08  &  1084 &   8.3 & PS outflows & 1,2,6 \\   
NGC\,6543         & AB & E, MS  &     48~~ & 4,600 &  9.4 &   1.50 & 0.07  &  613  &    43 & PS outflows & 1,7  \\
NGC\,6572         & AC & MP, MS &     69~~ &10,000 &  7.0 &   1.86 & 0.06  &  1107 &    32 & Outflow     & 1,8  \\    
NGC\,7009         & AB & E, MS  &     87~~ & 3,900 & 12.5 &   1.45 & 0.09  &  1270 &    22 & FLIERs      & 1,2  \\    
\hline
IC\,289           &  B & E, MS  &   $\dots$&   700 & 15.0 &   2.19 & 0.16  &   460 &     5 &             & 1,9  \\ 
IC\,2448          &  B & E, MS  &     65~~ & 1,100 & 11.0 &   2.20 & 0.12  &  1173 &   1.0 &             & 1,10 \\ 
NGC\,1501         &  B & E      &    135~~ &   950 & 27.0 &   0.72 & 0.09  &  1151 &  9.93 &             & 1,11 \\
NGC\,2371-2       &  B & E      &    100~~ & 2,500 & 28.0 &   0.50 & 0.07  &   901 &  160  & Ansae/Lobe  & 12   \\
NGC\,3242         &  B & E, MS  &     89~~ & 2,000 & 18.6 &   1.00 & 0.09  &  1520 &   14  & FLIERs      & 1,13 \\ 
NGC\,5307         &  B & E, MS  &   $\dots$& 2,500 &  7.1 &   3.0  & 0.10  &  1462 & 11.07 & PS outflows & 9,14  \\
NGC\,5882         &  B & E, MS  &     68~~ & 5,900 &  7.0 &   1.70 & 0.06  &  1335 & 15.5  &             & 1,15  \\ 
NGC\,6153         &  B & E, MS  &    109~~ & 5,100 & 12.5 &   1.10 & 0.07  &  1189 &  64.2 & FLIERs      & 1,16  \\
NGC\,6578         &  B & E, MS  &   $\dots$& 4,700 &  6.0 &   2.9  & 0.08  &   893 &   36  &             & 1,9  \\ 
NGC\,6818         &  B & E, MS  &    160~~ & 2,100 & 10.0 &   1.64 & 0.08  &  1190 &    46 &             & 1,17  \\
NGC\,6826         &  B & E, MS  &     50~~ & 3,400 & 12.7 &   1.30 & 0.08  &   734 &   9.3 & FLIERs      & 1,2  \\
NGC\,6884         &  B & PS, MS &     95~~ & 2,300 &  8.0 &   2.55 & 0.10  &  1966 &    26 & PS outflows & 1,18 \\ 
NGC\,6891         &  B & E, MS  &     50~~ & 3,000 &  6.6 &   2.9  & 0.09  &  868  &    12 &             & 1,9,19 \\
NGC\,7662         &  B & E, MS  &    111~~ & 2,700 & 14.8 &   1.19 & 0.08  &  1223 &     2 & FLIERs      & 1,20 \\
\hline
BD+30$^\circ$3639  &  C & E, MS  &     32~~ &40,800 &  3.0 &   1.3  & 0.02  &  9    &$\dots$& Collimated outflows & 1,9  \\ 
H\,1-9            &  C & E      &     35~~ &13,000 &  0.5 &   4.4  & 0.01  & 153.1 &  89.9 &                     & 21 \\ 
H\,1-46           &  C & PS     &     41~~ &$\dots$&  0.7 &   3.5  & 0.01  &   540 &   149 & Possible PS outflows& 22 \\
Hb\,12            &  C & B      &     35~~ &500,000&  5.5 &   10.0 & 0.27  &  509.3&   22.1&                     & 23,24 \\
Hen\,2-47         &  C &        &$\dots$~~ &$\dots$&  5.5 &   5.5  & 0.15  &  12   &   458 &                     & 9 \\
Hen\,2-260        &  C & E      &     27~~ &$\dots$&  0.5 &   6.9  & 0.02  &$\dots$&   213 & Ansae               & 9,22 \\   
Hen\,2-447        &  C & MP     &$\dots$~~ &$\dots$&  1.3 &   7.6  & 0.05  &  427  &708    &                     & 9,23\\   
Hen\,3-1357       &  C & B      &$\dots$~~ &$\dots$&  1.5 &   5.6  & 0.04  &$\dots$&$\dots$& Possible PS outflows& $-$ \\   
IC\,418           &  C & E, MS  &     38~~ &20,000 &  6.2 &   1.2  & 0.04  &  85.9 & 206.8 &                     & 1,25 \\ 
IC\,5117          &  C & MP     &     57~~ &38,900 &  0.7 &   1.3  & 0.005 &1346.4 &  47.9 &                     & 26 \\   
M\,1-19           &  C & E      &$\dots$~~ &$\dots$&  1.3 &   5.5  & 0.03  &   546 &   214 & Low SNR             & 22 \\
M\,1-20           &  C & E, MS  &     81~~ &11,200 &  0.8 &   3.4  & 0.01  &   830 &    78 &                     & 1,22 \\ 
M\,1-31           &  C & MP     &     90~~ &$\dots$&  3.5 &   4.2  & 0.07  &   737 &   516 & Possible PS outflows& 22 \\   
M\,1-42           &  C & E      &  $\dots$ &$\dots$&  7.8 & $\dots$&$\dots$&   590 &   246 & Ansae/bipolar lobes & 22 \\
M\,1-61           &  C & MP     &     73~~ &$\dots$&  1.8 &   2.4  & 0.02  &   914 &   170 & Possible PS outflows& 9 \\
M\,2-43           &  C & E      &     57~~ &$\dots$&  0.6 &   5.0  & 0.01  &   629 &  1162 & Ansae               & 9,23 \\   
MyCn\,18          &  C & B      &     51~~ &$\dots$&  7.2 &   3.1  & 0.11  &   294 &$\dots$& PS outflows         & $-$ \\ 
NGC\,2346         &  C & B      &    112~~ &   440 & 27.3 &   0.9  & 0.12  &  1001 &   395 &                     & 1,27 \\ 
NGC\,6565         &  C & E      &    120~~ & 3,200 &  5.0 &   2.25 & 0.06  &$\dots$&$\dots$& Ansae               & 1,28 \\ 
NGC\,6720         &  C & E, MS  &    116~~ &   800 & 35.0 &   0.7  & 0.12  &  1064 &   297 &                     & 1,9 \\ 
NGC\,6790         &  C & E, MS  &     74~~ & 9,600 &  2.5 &   1.5  & 0.02  &  1274 &    71 & Bipolar lobes       & 9 \\   
NGC\,6881         &  C & Q      &     45~~ & 7,300 &  2.5 &   2.5  & 0.03  &  1851 &   603 & H$_2$ hourglass lobes & 9 \\
NGC\,6886         &  C & B      &    129~~ & 4,900 &  3.8 &   5.3  & 0.10  &$\dots$&   346 &                     & 1,9 \\
NGC\,7027         &  C & B      &    175~~ &30,200 & 10.0 &   0.9  & 0.04  &$\dots$&$\dots$& PS outflows         & 1 \\ 
\hline
H\,1-8            &  D & MP     &$\dots$~~ &$\dots$&  1.1 &   5.3  & 0.03  &  993  &  2177 & Low SNR             & 22 \\   
H\,2-13           &  D & E, MS  &$\dots$~~ &$\dots$&  1.6 &$\dots$ &$\dots$&  2081 &   529 & Low SNR             & 22 \\   
H\,2-18           &  D & B      &$\dots$~~ &$\dots$&  1.9 &   7.8  & 0.07  &  1427 &    30 & Low SNR             & 22 \\   
H\,2-27           &  D & B      &$\dots$~~ &$\dots$&  1.8 &$\dots$ &$\dots$&   816 &   887 & Low SNR             & 22 \\   
H\,2-37           &  D & PS     &$\dots$~~ &$\dots$&  2.0 &$\dots$ &$\dots$&   811 &   198 & Low SNR             & 22 \\   
H\,2-39           &  D & $\dots$&$\dots$~~ &$\dots$&  2.2 &$\dots$ &$\dots$&  1509 &$\dots$& Low SNR             & 22 \\   
Hen\,2-90         &  D & PS     &$\dots$~~ &$\dots$&  2.3 &$\dots$ &$\dots$&   205 &   141 & Fast collimated outflows & 9 \\   
Hen\,3-1475       &  D & PS     &$\dots$~~ &$\dots$&  5.8 &$\dots$ &$\dots$&$\dots$&$\dots$& Fast collimated outflows & $-$\\   
KFL\,11           &  D & $\dots$&$\dots$~~ &$\dots$&  1.0 &$\dots$ &$\dots$&   910 &   127 & Low SNR             & 22 \\   
M\,1-92           &  D & B      &$\dots$~~ &$\dots$&  5.2 &$\dots$ &$\dots$&$\dots$&$\dots$&                     & $-$\\   
M\,2-9            &  D & Q      &     35~~ & 3,200 & 18.5 &   1.7  & 0.15  &    90 &   460 &                     & 9 \\   
M\,2-29           &  D & E, MS  &     68~~ &$\dots$&  2.0 &   8.6  & 0.08  &   516 &    34 & Low SNR             & 22 \\   
M\,3-14           &  D & B      &$\dots$~~ &$\dots$&  1.9 &   5.7  & 0.05  &  1065 &  1196 & Low SNR             & 22 \\   
M\,3-15           &  D & B      &$\dots$~~ & 4,200 &  2.0 &   5.3  & 0.05  &  1064 &   224 & Low SNR             & 22 \\   
M\,3-35           &  D & MP     &$\dots$~~ &$\dots$&  1.0 &   1.8  & 0.01  &  1123 &    34 &                     & 9 \\   
M\,3-38           &  D & B      &    153~~ &$\dots$&  0.7 &   5.9  & 0.02  &  1939 &   598 & Low SNR, probable Q & 9,22 \\   
MaC\,1-11         &  D & $\dots$&$\dots$~~ &$\dots$&  1.4 &$\dots$ &$\dots$&   995 &    49 & Low SNR             & 22 \\   
Necklace Nebula   &  D &        &$\dots$~~ &   820 & 13.5 &   4.6  & 0.30  &  534.6&  12.7 &                     & 29 \\
NGC\,2392         &  D & E, MS  &     47~~ &   900 & 23.0 &   1.28 & 0.14  &  1406 &   160 &                     & 1,9 \\ 
NGC\,5315         &  D &        &       ~~ &10,200 &  5.4 &   2.62 & 0.07  &   840 &   266 &                     & 1,9  \\
NGC\,6369         &  D & E, MS  &     66~~ & 3,800 & 16.0 &   1.55 & 0.12  &  1484 &   183 &                     & 1,9  \\
PM\,1-89          &  D & E      &$\dots$~~ &$\dots$&  3.7 &$\dots$ &$\dots$&$\dots$&$\dots$& Low SNR             & $-$\\   
Ps\,1             &  D & E, MS  &$\dots$~~ &$\dots$&  1.2 &   8.4  & 0.05  &   271 &     4 &                     & 9 \\   
StWr\,2-21        &  D & E      &$\dots$~~ &$\dots$&  0.9 &$\dots$ &$\dots$&  1065 &     8 & Low SNR             & 9 \\   
Th\,3-4           &  D & B      &$\dots$~~ &$\dots$&  0.9 &$\dots$ &$\dots$&  2005 &   821 & Low SNR             & 22 \\   
Wray\,16-286      &  D & Q      &$\dots$~~ &$\dots$&  1.1 &$\dots$ &$\dots$&  1259 &   419 & Low SNR, quadrupolar& 9 \\   
Wray\,16-423      &  D & $\dots$&$\dots$~~ &$\dots$&  0.8 &$\dots$ &$\dots$&  1198 &    19 &                     & 9 \\
\hline
\end{tabular}
\tablebib{
(1) \citet{Frew2008}; 
(2) \citet{KwitterHenry1998}; 
(3) \citet{Guerrero_etal2008}; 
(4) \citet{Hyung1999};
(5) \citet{Ercolano2003}; 
(6) \citet{Pottasch2009}; 
(7) \citet{WessonLiu2004}; 
(8) \citet{Hyung1994b}; 
(9) \citet{Acker1992}
(10) \citet{Guiles2007}; 
(11) \citet{Ercolano2004}; 
(12) \citet{Pottasch1981}; 
(13) \citet{Ruiz2011}; 
(14) \citet{Ruiz2003}; 
(15) \citet{Pottasch2004}; 
(16) \citet{Yuan2011}.
(17) \citet{Pottasch2005}; 
(18) \citet{HyungAller1997}; 
(19) \citet{Mendez1988}; 
(20) \citet{Hyung1997};
(21) \citet{Cavichia2010};
(22) \citet{Acker1991};
(23) \citet{Phillips2004};
(24) \citet{Hyung1996};
(25) \citet{Hyung1994b};
(26) \citet{Hyung2001};
(27) \citet{Sabbadin1976};
(28) \citet{Turatto2002};
(29) \citet{Corradi2011}
}
}
\end{table*}

\begin{table*}[!t]
\centering
\caption{
Quantitative properties of the {[O~{\sc iii}]/H$\alpha$} profiles 
of Figs.~\ref{profA}, \ref{profB}, and \ref{profC}.}
\begin{tabular}{llrrrr} 
\hline\hline
\multicolumn{1}{l}{PN} & 
\multicolumn{1}{l}{Region} & 
\multicolumn{1}{c}{[O~{\sc iii}]/H$\alpha$} &  
\multicolumn{3}{c}{\underline{~~~~~~~~~~~~~~~$FWHM$\tablefootmark{a}~~~~~~~~~~~~~~~}} \\ 
\multicolumn{1}{c}{} & 
\multicolumn{1}{c}{} & 
\multicolumn{1}{c}{~~~~~enhancement~~~~~} & 
\multicolumn{1}{c}{} & 
\multicolumn{1}{c}{} & 
\multicolumn{1}{c}{} \\ 
\multicolumn{1}{c}{} & 
\multicolumn{1}{c}{} & 
\multicolumn{1}{c}{} & 
\multicolumn{1}{c}{(\%)} & 
\multicolumn{1}{c}{(\arcsec)} & 
\multicolumn{1}{c}{(cm)} \\
\hline
\multicolumn{6}{c}{Type A PNe~~~~~~~~~~} \\
\hline
IC\,4593  & Southern bow-shock & $\times$2.0~~~~~~~~~~ &  2\%~ & 0\farcs27~  &  1.5$\times$10$^{16}$~ \\ 
NGC\,6210 & Northern bow-shock & $\times$1.7~~~~~~~~~~ &  3\%~ & 0\farcs70~  &  2.1$\times$10$^{16}$~ \\ 
NGC\,7009 & Eastern bow-shock  & $\times$1.3~~~~~~~~~~ &  3\%~ & 0\farcs63~  &  1.1$\times$10$^{16}$~ \\ 
\hline
\multicolumn{6}{c}{Type B PNe~~~~~~~~~~} \\
\hline
NGC\,3242 & Inner shell        & $\times$1.3~~~~~~~~~~ & 10\%~ & 0\farcs93~  &  3.6$\times$10$^{16}$~ \\ 
          & Outer shell        & $\times$2.7~~~~~~~~~~ &  3\%~ & 0\farcs57~  &  2.2$\times$10$^{16}$~ \\ 
          & North-west blister & $\times$2.0~~~~~~~~~~ & 1.5\% & 0\farcs38~  &  1.5$\times$10$^{16}$~ \\ 
NGC\,6826 & Outer shell        & $\times$3.1~~~~~~~~~~ &  1\%~ & 0\farcs14~  &  4.0$\times$10$^{15}$~ \\ 
\hline
\multicolumn{6}{c}{Type C PNe~~~~~~~~~~} \\
\hline
IC\,418   & Main shell         &   $\div$6.0~~~~~~~~~~ & 40\%~ & 1\farcs7~~~ &  1.5$\times$10$^{16}$~ \\ 
NGC\,6720 & Main shell         &   $\div$4.9~~~~~~~~~~ & 24\%~ & 7\farcs5~~~ & 1.0$\times$10$^{17}$~ \\ 
NGC\,6886 & Main shell         &   $\div$1.7~~~~~~~~~~ & 18\%~ & 0\farcs32~  &  1.5$\times$10$^{16}$~ \\ 
\hline
\end{tabular}
\tablefoot{
\tablefoottext{a}{
The shape of the profile of these regions is not symmetric and the 
measurement of their widths corresponds to the innermost half of the 
profile.  
}}
\label{tab3}

\end{table*}


\begin{thebibliography}{}

\bibitem[Acker et al.(1991)]{Acker1991} Acker, A., Raytchev, B.,
  Koeppen, J., \& Stenholm, B.\ 1991, \aaps, 89, 237

\bibitem[Acker et al.(1992)]{Acker1992} Acker, A., Marcout, J.,
  Ochsenbein, F., et al.\ 1992, The Strasbourg-ESO Catalogue of
  Galactic Planetary Nebulae.~Parts I, II., by Acker, A.; Marcout, J.;
  Ochsenbein, F.; Stenholm, B.; Tylenda, R.; Schohn, C..~ European
  Southern Observatory, Garching (Germany), 1992, 1047 p., ISBN
  3-923524-41-2

\bibitem[B\"{a}ssgen et al.(1990)]{Bassgen1990} B\"{a}ssgen, M., Diesch,
  C., \& Grewing, M.\ 1990, \aap, 237, 201

\bibitem[Balick(2004)]{Balick2004} 
Balick, B.\ 2004, 
\aj, 127, 2262

\bibitem[Balick \& Frank(2002)]{BF02} 
Balick, B., \& Frank, A.\ 2002, 
ARA\&A, 40, 439

\bibitem[Bohigas(2001)]{Bohigas2001} Bohigas, J.\ 2001, \rmxaa, 37,
  237

\bibitem[Cavichia et al.(2010)]{Cavichia2010} Cavichia, O., Costa,
  R.~D.~D., \& Maciel, W.~J.\ 2010, \rmxaa, 46, 159

\bibitem[Corradi et al.(1997a)]{Corradi_etal1997} 
Corradi, R.~L.~M., Guerrero, M., Manchado, A., \& Mampaso, A.\ 1997a, 
\na, 2, 461 

\bibitem[Corradi et al.(1996)]{Corradi_etal96} 
Corradi, R.~L.~M., Manso, R., Mampaso, A., \& Schwarz, H.~E.\ 1996, 
\aap, 313, 913 

\bibitem[Corradi et al.(1997b)]{Corradi1997} 
Corradi, R.~L.~M., Perinotto, M., Schwarz, H.~E., \& Claeskens, J.-F.\ 1997b, 
\aap, 322, 975

\bibitem[Corradi et al.(2011)]{Corradi2011} Corradi, R.~L.~M., Sabin,
  L., Miszalski, B., et al.\ 2011, \mnras, 410, 1349

\bibitem[Cox(1972)]{Cox1972} 
Cox, D.~P.\ 1972, 
\apj, 178, 143 

\bibitem[Dennis et al.(2008)]{Dennis_etal08}
Dennis, T.~J., Cunningham, A.~J., Frank, A., et al.\ 2008, 
\apj, 679, 1327

\bibitem[Ercolano et al.(2003)]{Ercolano2003} Ercolano, B., Morisset,
  C., Barlow, M.~J., Storey, P.~J., \& Liu, X.-W.\ 2003, \mnras, 340,
  1153

\bibitem[Ercolano et al.(2004)]{Ercolano2004} Ercolano, B., Wesson,
  R., Zhang, Y., et al.\ 2004, \mnras, 354, 558

\bibitem[Frew(2008)]{Frew2008}
Frew, D.\ 2008, PhD Thesis, Macquarie University, Sydney, Australia

\bibitem[Gruendl et al.(2000)]{Gruendl_etal2000} 
Gruendl, R.~A., Chu, Y.-H., Dunne, B.~C., \& Points, S.~D.\ 2000, 
\aj, 120, 2670 

\bibitem[Guerrero et al.(2008)]{Guerrero_etal2008} 
Guerrero, M.~A., et al.\ 2008, 
\apj, 683, 272 

\bibitem[Guiles et al.(2007)]{Guiles2007} Guiles, S., Bernard-Salas,
  J., Pottasch, S.~R., \& Roellig, T.~L.\ 2007, \apj, 660, 1282

\bibitem[Hartigan et al.(1987)]{HRH87} 
Hartigan, P., Raymond, J., \& Hartmann, L.\ 1987, 
\apj, 316, 323 

\bibitem[Hyung \& Aller(1996)]{Hyung1996} 
Hyung, S., \& Aller, L.~H.\ 1996, 
\mnras, 278, 551

\bibitem[Hyung \& Aller(1997)]{HyungAller1997} 
Hyung, S., \& Aller, L.~H.\ 1997, \apj, 491, 242

\bibitem[Hyung et al.(1994a)]{Hyung1994a} 
Hyung, S., Aller, L.~H., \& Feibelman, W.~A.\ 1994, \mnras, 269, 975

\bibitem[Hyung et al.(1994b)]{Hyung1994b} 
Hyung, S., Aller, L.~H., \& Feibelman, W.~A.\ 1994, \pasp, 106, 745

\bibitem[Hyung et al.(1997)]{Hyung1997} 
Hyung, S., Aller, L.~H., \& Feibelman, W.~A.\ 1997, \apjs, 108, 503

\bibitem[Hyung et al.(1999)]{Hyung1999} Hyung, S., Aller, L.~H., \&
  Feibelman, W.~A.\ 1999, \apj, 525, 294

\bibitem[Hyung et al.(2001)]{Hyung2001} Hyung, S., Aller, L.~H.,
  Feibelman, W.~A., \& Lee, S.-J.\ 2001, \apj, 563, 889

\bibitem[Kwitter \& Henry(1998)]{KwitterHenry1998} Kwitter, K.~B., \&
  Henry, R.~B.~C.\ 1998, \apj, 493, 247


\bibitem[Lee \& Sahai(2003)]{LS03}
Lee, C.F., \& Sahai, R.\ 2003, 
\apj, 586, 319.

\bibitem[Lim et al.(2010)]{Lim_etal2010}
Lim, P.L.\ et al.\ 2010, Space Telescope Science Institute, Instrument 
Science Report WFPC2 2010-05

\bibitem[M\'endez et al.(1988)]{Mendez1988} 
M\'endez, R.~H., Kudritzki, R.~P., Herrero, A., Husfeld, D., \& Groth, 
H.~G.\ 1988, \aap, 190, 113

\bibitem[M\'endez \& Niemela(1981)]{MN81} 
M\'endez, R.~H., \& Niemela, V.~S.\ 1981, \apj, 250, 240 

\bibitem[Moore et al.(2002)]{Moore_etal2002} 
Moore, B.~D., Walter, D.~K., Hester, J.~J., et al.\ 2002, \aj, 124, 3313 

\bibitem[Napiwotzki(1999)]{Napiwotzki99} 
Napiwotzki, R.\ 1999, \aap, 350, 101 

\bibitem[O'Dell et al.(2013)]{ODell_etal2013} 
O'Dell, C.~R., Ferland, G.~J., Henney, W.~J., \& Peimbert, M.\ 2013, 
\aj, 145, 92 

\bibitem[Perinotto et al.(2004)]{Petal04} 
Perinotto, M., Sch{\"o}nberner, D., Steffen, M., \& Calonaci, C.\ 2004, 
\aap, 414, 993 

\bibitem[Phillips \& Cuesta(1998)]{Phillips1998} 
Phillips, J.~P., \& Cuesta, L.\ 1998, \aaps, 133, 381

\bibitem[Phillips(2004)]{Phillips2004} 
Phillips, J.~P.\ 2004, \mnras, 353, 589

\bibitem[Pottasch et al.(1981)]{Pottasch1981} 
Pottasch, S.~R., Gathier, R., Gilra, D.~P., \& Wesselius, P.~R.\ 1981, 
\aap, 102, 237

\bibitem[Pottasch et al.(2004)]{Pottasch2004} 
Pottasch, S.~R., Bernard-Salas, J., Beintema, D.~A., \& Feibelman, W.~A.\ 2004, 
\aap, 423, 593

\bibitem[Pottasch et al.(2005)]{Pottasch2005} 
Pottasch, S.~R., Beintema, D.~A., \& Feibelman, W.~A.\ 2005, \aap, 436, 953

\bibitem[Pottasch et al.(2009)]{Pottasch2009} 
Pottasch, S.~R., Bernard-Salas, J., \& Roellig, T.~L.\ 2009, \aap, 499, 249

\bibitem[Pottasch \& Surendiranath(2005)]{PS05} 
Pottasch, S.~R., \& Surendiranath, R.\ 2005, \aap, 432, 139 


\bibitem[Raga et al.(2008)]{Raga_etal08}
Raga, A.C., Riera, A., Mellema, G., Esquivel, A., \& Vel\'azquez, P.F.\ 2008, 
\aap, 489, 1141

\bibitem[Raga et al.(2002)]{Raga_etal02}
Raga, A.C., Vel\'azquez, P.F., Cant\'o, J., \& Masciadri, E.\ 2002, 
\aap, 395, 647 

\bibitem[Ramos-Larios et al.(2012)]{R-L_etal2012} 
Ramos-Larios, G., V{\'a}zquez, R., Guerrero, M.~A., et al.\ 2012, 
\mnras, 423, 3753 

\bibitem[Reipurth et al.(1997)]{Reipurth_etal1997} 
Reipurth, B., Hartigan, P., Heathcote, S., Morse, J.~A., \& Bally, J.\ 1997, 
\aj, 114, 757 

\bibitem[Reipurth et al.(2002)]{Reipurth_etal2002} 
Reipurth, B., Heathcote, S., Morse, J., Hartigan, P., \& Bally, J.\ 2002, 
\aj, 123, 362 

\bibitem[Rodr{\'{\i}}guez-Gonz{\'a}lez et al.(2012)]{R-Getal12} 
Rodr\'\i guez-Gonz\'alez, A., Esquivel, A., Raga, A.~C., et al.\ 2012, 
\aj, 143, 60 

\bibitem[Ruiz et al.(2003)]{Ruiz2003} 
Ruiz, M.~T., Peimbert, A., Peimbert, M., \& Esteban, C.\ 2003, 
\apj, 595, 247

\bibitem[Ruiz et al.(2011)]{Ruiz2011} 
Ruiz, N., Guerrero, M.~A., Chu, Y.-H., \& Gruendl, R.~A.\ 2011, 
\aj, 142, 91

\bibitem[Sabbadin(1976)]{Sabbadin1976} 
Sabbadin, F.\ 1976, \aap, 52, 291

\bibitem[Sahai \& Trauger(1998)]{ST98}
Sahai, R., \& Trauger, J.T.\ 1998, 
\aj, 116, 1357.

\bibitem[Sch{\"o}nberner et al.(2005)]{Setal05} 
Sch{\"o}nberner, D., Jacob, R., Steffen, M., et al.\ 2005, 
\aap, 431, 963 

\bibitem[Turatto et al.(2002)]{Turatto2002} Turatto, M., Cappellaro,
  E., Ragazzoni, R., Benetti, S., \& Sabbadin, F.\ 2002, \aap, 384,
  1062

\bibitem[V{\'a}zquez(2012)]{Vazquez2012} 
V\'azquez, R.\ 2012, 
\apj, 751, 116


\bibitem[Villaver et al.(2002)]{Villaver_etal2002} 
Villaver, E., Manchado, A., \& Garc{\'{\i}}a-Segura, G.\ 2002, 
\apj, 581, 1204 

\bibitem[Yuan et al.(2011)]{Yuan2011} Yuan, H.-B., Liu, X.-W.,
  P{\'e}quignot, D., et al.\ 2011, \mnras, 411, 1035

\bibitem[Walton et al.(1986)]{Walton1986} Walton, N.~A., Reay, N.~K.,
  Pottasch, S.~R., \& Atherton, P.~D.\ 1986, New Insights in
  Astrophysics.~ Eight Years of UV Astronomy with IUE, 263, 497

\bibitem[Wesson \& Liu(2004)]{WessonLiu2004} Wesson, R., \& Liu,
  X.-W.\ 2004, \mnras, 351, 1026

\end{thebibliography}
\end{document}